\documentclass[10pt]{article}
\usepackage{epsfig,epsf}

\usepackage{comment}

\usepackage{amsmath}
\usepackage{amsthm}
\usepackage{amsfonts}
\usepackage{amssymb}
\usepackage{cite}
\usepackage{slashed}
\usepackage{amscd}
\usepackage{graphicx}
\renewcommand{\theequation}{\arabic{section}.\arabic{equation}}
\renewcommand{\thefootnote}{\fnsymbol{footnote}}
\renewcommand{\thetable}{\arabic{table}}

\renewcommand{\theequation}{\arabic{section}.\arabic{equation}}
\renewcommand{\thefootnote}{\fnsymbol{footnote}}
\renewcommand{\thetable}{\arabic{table}}

\def\be{\begin{equation}}
\def\ee{\end{equation}}
\def\lb{\label}

\newcommand{\as}{\ifmmode\alpha_{\rm s}\else{$\alpha_{\rm s}$}\fi}
\newcommand{\asbar}{\ifmmode\bar{\alpha}_{\rm s}\else{$\bar{\alpha}_{\rm s}$}\fi}

\font\cmss=cmss12 
\def\inbar{\,\vrule height1.5ex width.4pt depth0pt}
\def\IC{\relax\hbox{$\inbar\kern-.3em{\rm C}$}}
\def\IZ{\relax{\hbox{\cmss Z\kern-.4em Z}}}
\def\IR{{\hbox{{\rm I}\kern-.2em\hbox{\rm R}}}}

\def\IP{{\hbox{{\rm I}\kern-.2em\hbox{\rm P}}}}
\def\II{\hbox{{1}\kern-.25em\hbox{l}}}

\newbox\lett\newdimen\lheight\newdimen\lwidth
\def\ontop#1#2{\setbox\lett=\hbox{#2}\lheight\ht\lett
\multiply\lheight by 12 \divide\lheight by 10\relax%
\lwidth\wd\lett \multiply\lwidth by 8 \divide\lwidth by 10\relax #2\kern-\lwidth%
\raise\lheight\hbox{{$\scriptstyle #1$}}\kern.1ex}

\def\inbar{\,\vrule height1.5ex width.4pt depth0pt}

\begin{document}


\begin{titlepage}

\vspace*{1cm}

\begin{center}
{\Large \bf{Spinorial R-matrix}}

\vspace{1cm}

{\large \sf D. Chicherin$^{ca}$\footnote{\sc e-mail: chicherin@pdmi.ras.ru},
  S. Derkachov$^{a}$\footnote{\sc e-mail: derkach@pdmi.ras.ru} and  A.P.
Isaev$^b$\footnote{\sc e-mail: isaevap@theor.jinr.ru} \\
}

\vspace{0.5cm}

\begin{itemize}
\item[$^a$]
{\it St. Petersburg Department of Steklov Mathematical Institute
of Russian Academy of Sciences,
Fontanka 27, 191023 St. Petersburg, Russia}
\item[$^b$]
{\it Bogoliubov Laboratory of Theoretical Physics,
JINR, 141 980 Dubna, Moscow region,  \\
and ITPM, M.V.Lomonosov Moscow State University, Russia}
\item[$^c$]
{\it Chebyshev Laboratory, St.-Petersburg State University,\\
14th Line, 29b, Saint-Petersburg, 199178 Russia}
\end{itemize}
\end{center}

\vspace{0.5cm}

\begin{abstract}
R-matrix acting in the tensor product of two spinor representation spaces
of Lie algebra so(d) is considered thoroughly. Corresponding
Yang-Baxter equation is proved. The relation to the local Yang-Baxter
relation is established.
\end{abstract}

\vspace{1cm}



{\small \tableofcontents}
\renewcommand{\refname}{References}
\renewcommand{\thefootnote}{\arabic{footnote}}
\setcounter{footnote}{0} \setcounter{equation}{0}

\end{titlepage}

\section{Introduction}

In this paper we are going to prove certain relations concerning
Yangian for $so(N)$ which has been formulated in \cite{CDI} and
to consider thoroughly corresponding numerical $\mathrm{R}$-matrix defining
Yangian for $so(N) \simeq {\sf spin}(N)$.

Let ${\cal A}$ be a Lie algebra and
 $T_a$ $(a=0,1,2,\dots)$ be representations of ${\cal A}$ in spaces $V_a$.
 Consider operators $\mathrm{R}_{ab}(u) \in {\rm End}(V_a \otimes V_b)$, where $u$ denotes a spectral
 parameter. We say that ${\cal A}$ is the symmetry algebra of the operator $\mathrm{R}_{ab}(u)$
 if $\forall g \in {\cal A}$ we have
 $$
 \Bigl(T_a(g) \otimes I_b + I_a \otimes T_b(g) \Bigr) \, \mathrm{R}_{ab}(u) =
 \mathrm{R}_{ab}(u)  \,  \Bigl(T_a(g) \otimes I_b + I_a \otimes T_b(g) \Bigr) \; ,
 $$
 where $I_a$ and $I_b$ are unit operators in $V_a$ and $V_b$, respectively.
 Consider a set of Yang-Baxter $\mathrm{RRR}$-equations for the operators $\mathrm{R}_{ab}(u)$:
\begin{equation} \label{RRR}
 \mathrm{R}_{ab}(u - v) \, \mathrm{R}_{bc}(u) \,  \mathrm{R}_{ab}(v) =
  \mathrm{R}_{bc}(v) \, \mathrm{R}_{ab}(u)  \,  \mathrm{R}_{bc}(u - v) \;\; \in  \;\;
  {\rm End}(V_a \otimes V_b \otimes V_c)
\end{equation}
where the representation spaces $V_a,\, V_b,\, V_c$ are different in general situation.
There is an efficient procedure which enables us to construct nontrivial solutions of cubic Yang-Baxter
 equations (\ref{RRR}) starting with the known one.
The procedure can be illustrated by the following sequence of specializations
in the Yang-Baxter relations (\ref{RRR}):
$$
V_{0}\otimes V_{0}\otimes V_{0} \to V_{a}\otimes V_{0}\otimes V_{0} \to
V_{a}\otimes V_{a}\otimes V_{0} \to V_{a}\otimes V_{a}\otimes V_{a} \to
V_{b}\otimes V_{a}\otimes V_{a} \to V_{b}\otimes V_{b}\otimes V_{a} \to V_{b}\otimes V_{b}\otimes V_{b} \to \cdots
$$
and corresponding sequence of solutions
$$
\mathrm{R}_{0,0} \to \mathrm{R}_{a,0} \to \mathrm{R}_{a,a} \to \mathrm{R}_{b,a} \to \mathrm{R}_{b,b} \to \cdots
$$
Indeed, one starts with the simplest known solution $\mathrm{R}_{0,0}$ of the Yang-Baxter equation (\ref{RRR})
defined in the space $V_{0}\otimes V_{0}\otimes V_{0}$, where
$V_{0}$ is the space of the simplest faithful representation, e.g. {\it defining} representation
for the matrix Lie algebra ${\cal A}$.
Further one introduces another representation $T_a$ which acts in the space $V_a$
(finite-dimensional or infinite-dimensional) and solves the Yang-Baxter equation (\ref{RRR})
restricted to $V_{a} \otimes V_{0} \otimes V_{0}$.
It happens to be a quadratic equation on the operator $\mathrm{R}_{a,0}$ which in special cases represents
 the Yangian of the
corresponding matrix Lie algebra ${\cal A}$ ($V_0$ is the space of the defining representation
and $V_a$ is the space of the representation of the Yangian).
On the next step one solves Yang-Baxter relation (\ref{RRR}) restricted to
the space $V_{a}\otimes V_{a}\otimes V_{0}$ and obtains $\mathrm{R}_{a,a}$.
There is a well known argumentation (based on the associativity ideas)
why $\mathrm{R}_{a,a}$ respects Yang-Baxter equation (\ref{RRR})
defined in the space $V_{a}\otimes V_{a}\otimes V_{a}$.
Nevertheless it can be proved directly. Thus, the solution $\mathrm{R}_{a,a}$ of the cubic Yang-Baxter equation
is constructed in several steps starting with the simplest one $\mathrm{R}_{0,0}$, and in each step linear or quadratic relations have to be solved.

To be more concrete let us remind \cite{CDI,Witten} how it works for
the algebra ${\cal A} \simeq so(d) \simeq {\sf spin}(d)$, where we assume $d$ to be {\it even}.
All the following formulae can be rewritten straightforwardly for $so(p,q)$ as well where $p+q = d$.
Corresponding {\it fundamental} $\mathrm{R}$-matrix $\mathrm{R}^{0}(u)$ defined in the tensor product
$V_{0}\otimes V_{0}$
of two fundamental (defining) $d$-dimensional representations of $so(d)$
can be represented as
\begin{equation} \label{Rfund1}
 (\mathrm{R}^{0})^{i_1 i_2}_{j_1 j_2}(u) = u \, \delta^{i_1}_{j_2} \delta^{i_2}_{j_1} +
 \delta^{i_1}_{j_1} \delta^{i_2}_{j_2} - \frac{u}{u+\frac{d}{2}-1} \delta^{i_1 i_2} \delta_{j_1 j_2} \; ,
\end{equation}
and depicted as follows
\be \label{Rfund}
\mathrm{R}^{0}(u) = u \begin{array}{c} \includegraphics[width=1.0
cm,angle=90]{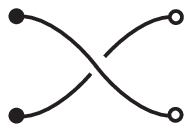} \end{array} +
\begin{array}{c} \includegraphics[width=1.0 cm,angle=90]{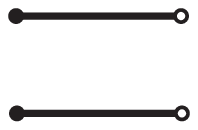} \end{array}
- \frac{u}{u+\frac{d}{2}-1} \begin{array}{c} \includegraphics[width=1.0
cm,angle=90]{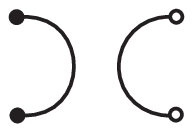} \end{array}
\ee
$\mathrm{R}^{0}$ respects Yang-Baxter equation
\begin{equation}
\mathrm{R}^{0}_{12}(u - v) \, \mathrm{R}^{0}_{23}(u) \,  \mathrm{R}^{0}_{12}(v) =
\mathrm{R}^{0}_{23}(v) \, \mathrm{R}^{0}_{12}(u)  \,  \mathrm{R}^{0}_{23}(u - v)
\;\; \in  \;\; {\rm End}(V_{0} \otimes V_{0} \otimes V_{0})
\end{equation}
and can be considered as the simplest solution in the hierarchy of solutions of the universal Yang-Baxter equation (\ref{RRR})
related to $so(d)$ Lie algebra.
It has been introduced in 1978 by A.~Zamolodchikov and Al.~Zamolodchikov \cite{Zam}.

On the next step we introduce {\it spinor} representation
of $so(d)$ acting in the space $V$ with dimension $2^{\frac{d}{2}}$.
Let $\gamma_a$ $(a=1,\dots,d)$
be $2^{\frac{d}{2}}$-dimensional
gamma-matrices in $\mathbb{R}^{d}$ which act in $V$ as linear operators. Operators $\gamma_{a}$ represent
generators of the Clifford algebra
\begin{equation} \lb{def}
\gamma_a \, \gamma_b  + \gamma_b \, \gamma_a = 2 \, \delta_{ab} \cdot \mathbf{1} \; .
\end{equation}
As a vector space, the Clifford algebra has dimension $2^{d}$.
The standard basis in this
space is formed by antisymmetrized products
of the $\gamma$-matrices
\begin{equation}
\label{Gamma01}
\gamma_{a_1\ldots a_k} = \frac{1}{k!}
\sum_{s }(-1)^{\mathrm{p}(s)} \gamma_{s(a_1)} \cdots\gamma_{s(a_k)} \equiv \gamma_{A_k}
\;\;\;\;\; (\forall  k \leq d) \; , \;\;\;\;
\gamma_{A_k} = 0 \;\;\;\;\; (\forall  k > d) \; ,
\end{equation}
where the summation is taken over all permutations $s$ of $k$ indices
$\{a_1, \dots, a_k\} \to \{s(a_1), \dots, s(a_k)\}$ and
$\mathrm{p}(s)$ denote the parity of the permutation $s$, $A_k$ is a multi-index $a_1\ldots a_k$.

Then, according to procedure outlined above, we look for the operator
 $\mathrm{L}^{0}(u)$ defined in the space $V_{0}\otimes V$ which respects quadratic relation
$$
 \mathrm{R}^{0}_{23}(u - v) \, \mathrm{L}^{0}_{12}(u) \,  \mathrm{L}^{0}_{13}(v) =
  \mathrm{L}^{0}_{12}(v) \, \mathrm{L}^{0}_{13}(u)  \,  \mathrm{R}^{0}_{23}(u - v)
   \;\;\in\;\;  {\rm End}(V \otimes V_{0} \otimes V_{0}) \; ,
$$
 where $\mathrm{R}^{0}_{23}(u)$ is fundamental $\mathrm{R}$-matrix (\ref{Rfund1}).
The solution of the above equation has been found in \cite{Witten}
 (see also \cite{KarT,ZamL,Resh}). It has the form
  \begin{equation}
 \lb{Lbox}
\mathrm{L}^{0}(u) =u \, \mathbf{1} \otimes I_n  - \frac{1}{4}\, [ \gamma^{a} , \gamma^{b} ]\otimes\,e_{ab}
 \end{equation}
where $e_{ab}$ are matrix units,
$\mathbf{1}$ and $I_n$ are identity operators
in spinor and defining representation spaces respectively,
summation over repeated indices is implied.

Further from the universal Yang-Baxter equation (\ref{RRR}) we obtain
a linear equation for $\mathrm{R}$-matrix $\mathrm{R}_{12}(u)$ acting in the tensor product $V\otimes V$ of two
spinor representations
\begin{equation} \label{RLLssf}
\mathrm{R}_{12}(u - v) \, \mathrm{L}^{0}_{13}(u) \,  \mathrm{L}^{0}_{23}(v) =
\mathrm{L}^{0}_{13}(v) \, \mathrm{L}^{0}_{23}(u)  \,  \mathrm{R}_{12}(u - v)
   \;\; \in  \;\; {\rm End}(V \otimes V \otimes V_{0})\,.
\end{equation}
In \cite{Witten} (see also \cite{KarT,Resh,ZamL})
spinorial $\mathrm{R}$-matrix has been sought for in $SO(d)$-invariant form
\begin{equation}
\label{r-mtr01}
\mathrm{R}(u) = \sum_{k=0}^{\infty} \frac{\mathrm{R}_k(u)}{k!}\, \gamma_{a_1\ldots a_k} \otimes
 \gamma^{a_1\ldots a_k} \;\;
  \in  \;\; {\rm End}(V \otimes V ) \; .
\end{equation}
For convenience, in the r.h.s. of (\ref{r-mtr01}),
the summation over $k$ runs up to infinity. However we note that this summation is
automatically truncated due to condition $k \leq d$ (see (\ref{Gamma01})).
It has been claimed in \cite{Witten} that
the $\mathrm{R}$-matrix (\ref{r-mtr01}) satisfies $\mathrm{RLL}$-relation (\ref{RLLssf})
if coefficient functions $\mathrm{R}_k(u)$
obey the recurrent relation
\begin{equation} \label{recurr}
\mathrm{R}_{k+2}(u) = \frac{u+k}{k-(u+d-2)} \,\mathrm{R}_{k}(u)\;.
\end{equation}
As far as we know it has not been checked directly so far that
$\mathrm{R}(u)$ satisfies Yang-Baxter relation defined in the space $V \otimes V \otimes V$
\begin{equation} \label{YB}
\mathrm{R}_{12}(u) \, \mathrm{R}_{23}(u+v) \, \mathrm{R}_{12}(v) =
\mathrm{R}_{23}(v) \, \mathrm{R}_{12}(u+v) \, \mathrm{R}_{23}(u)
\;\; \in  \;\; {\rm End}(V \otimes V \otimes V)
\end{equation}
owing to complicated gamma-matrix structure one has to deal with.
One of the aims of this paper is to carry out corresponding calculation.
In order to avoid multiple summation over repeated indices
we apply the generating functions technique and rewrite the sum in (\ref{r-mtr01})
as an integral over auxiliary parameter.
It enables us to perform the calculation in a concise manner.
We undertake this calculation in Section \ref{Yang-Baxter}.
The derivation of the recurrent relations~(\ref{recurr}) is given in Appendix.

Now we proceed to explore thoroughly spinorial $\mathrm{R}$-matrix (\ref{r-mtr01}).
At first we are going to deduce its basic properties
which can be obtained on a rather general argumentation
and do not need intricate calculation technique.
Further we formulate the other properties
which we prove in subsequent Sections.

It is well known that spinor representation $T$ for generators $M_{ab}$
of the Lie algebra $so(d)$ can be
constructed out of gamma-matrices $T(M_{ab}) = \frac{i}{2} \gamma_{ab}$ (\ref{Gamma01}).
Then one can easily check that
$\mathrm{R}$-matrix (\ref{r-mtr01}) is invariant under ${\sf spin}(d)$ action, i.e.
$$
(\gamma_{ab} \otimes \mathbf{1} + \mathbf{1} \otimes \gamma_{ab}) \; \mathrm{R}(u) =
\mathrm{R}(u) \; (\gamma_{ab} \otimes \mathbf{1} + \mathbf{1} \otimes \gamma_{ab})  \; ,  \;\;\;\;
\forall a,b \; ,
$$
moreover it satisfies commutation relation
\begin{equation} \label{Gam}
(\gamma_{d+1} \otimes \gamma_{d+1}) \; \mathrm{R}(u)
= \mathrm{R}(u) \; (\gamma_{d+1} \otimes \gamma_{d+1})\, ,
\end{equation}
which demonstrates an additional $u(1)$ symmetry of $\mathrm{R}(u)$.
In (\ref{Gam}) matrix $\gamma_{d+1}$ is defined as follows
\be \lb{chiral}
\gamma_{d+1} = \alpha \, \gamma_{1\cdots d} \;\;,\;\; \alpha^2 = (-)^{\frac{d}{2}} \;\;;\;\;
\gamma_{d+1}^2 = \mathbf{1} \;\; ; \;\;\; \{ \gamma_{d+1} \,,\, \gamma_a \}= 0 \;\;
\text{at} \;\; a = 1 , \cdots , d
\ee
and in the appropriate representation
of gamma-matrices it takes the form
$\gamma_{d+1} = \mathrm{diag} (I,-I)$.

Let us note that recurrence equations (\ref{recurr})
for the series of even coefficient functions $\mathrm{R}_{2k}(u)$
and odd ones $\mathrm{R}_{2k+1}(u)$ are independent.
The general solutions of these equations are
\begin{equation} \label{Revenodd}
\mathrm{R}_{2k}(u) = \mathrm{A}(u) \, (-1)^k  \,\frac{\Gamma(k+\frac{u}{2})\Gamma(\frac{u+d}{2}-k)}
{\Gamma(\frac{u}{2})\Gamma(\frac{u}{2})} \;,\;\;\;
\mathrm{R}_{2k+1}(u) =  \mathrm{B}(u) \, (-1)^k  \,\frac{\Gamma(k+\frac{u+1}{2})
\Gamma(\frac{u+d-1}{2}-k)}{2\,\Gamma(\frac{u+1}{2})\Gamma(\frac{u+1}{2})}
\end{equation}
where $\mathrm{A}$ and $\mathrm{B}$ are arbitrary functions of spectral parameter $u$.
For example, if $\mathrm{A}$ and $\mathrm{B}$ are polynomials  of spectral parameter
then coefficient functions in (\ref{Revenodd}) are normalized to be polynomials as well.
Thus it is convenient to decompose spinorial $\mathrm{R}$-matrix (\ref{r-mtr01}) in the sum
$\mathrm{R}(u) = \mathrm{R}^{+}(u) + \mathrm{R}^{-}(u)$
where (\ref{Gamma01})
\be \lb{Rpm}
\mathrm{R}^{+}(u) = \sum_{k=0}^{\infty}\frac{\mathrm{R}_{2k}(u)}{(2k)!}\,
\gamma_{A_{2k}} \otimes \gamma^{A_{2k}}
\;\; , \;\; \mathrm{R}^{-}(u) =
\sum_{k=0}^{\infty}\frac{\mathrm{R}_{2k+1}(u)}{(2k+1)!}\, \gamma_{A_{2k+1}}
\otimes \gamma^{A_{2k+1}}\;.
\ee
We refer to $\mathrm{R}^{+}(u)$ and $\mathrm{R}^{-}(u)$ as {\it even} and {\it odd} parts
of spinorial $\mathrm{R}$-matrix, respectively.

Consider the decomposition of the  spinorial $\mathrm{R}$-matrix in the sum
\begin{equation} \lb{decomp1}
\mathrm{R}(u) = \mathrm{P}^{+} \, \mathrm{R}(u)  + \mathrm{P}^{-} \, \mathrm{R}(u) \; ,
\end{equation}
where $\mathrm{P}^{\pm}$ are {\it projectors}
\begin{equation}
\lb{proj02}
\mathrm{P}^{\pm}  = \frac{1}{2} \left( \mathbf{1} \otimes \mathbf{1} \pm  \gamma_{d+1} \otimes \gamma_{d+1} \right) \;;\;\;
\mathrm{P}^{+}  \mathrm{P}^{-} = \mathrm{P}^{-}  \mathrm{P}^{+} = 0 \;,\;\;\; (\mathrm{P}^{\pm})^2  = \mathrm{P}^{\pm}\,.
\end{equation}

\noindent
{\bf Proposition 1.} Even and odd parts  (\ref{Rpm}) of the spinorial $\mathrm{R}$-matrix can
be singled out by projectors $\mathrm{P}^{\pm}$, i.e. we have
 \be
 \lb{decomp5}
 \;\;\;
\mathrm{R}^{+}(u) = \mathrm{P}^{+} \, \mathrm{R}(u) \; , \;\;\;
\mathrm{R}^{-}(u) = \mathrm{P}^{-} \, \mathrm{R}(u) \; ,
 \ee
\begin{equation}
\lb{proj01}
\mathrm{P}^{\pm} \, \mathrm{R}^{\pm}(u) = \mathrm{R}^{\pm}(u) \;
,\;\;\; \mathrm{P}^{\pm} \, \mathrm{R}^{\mp}(u) = 0 \;,\;\;\;
\mathrm{R}^{\pm}(u) \,\mathrm{R}^{\mp}(v) = 0 \; .
\end{equation}
{\bf Proof.} Due to (\ref{Revenodd}) we see that
coefficient functions in (\ref{Rpm}) satisfy the reciprocal conditions
\be
 \lb{decomp6}
\mathrm{R}_{2k}(u) = (-1)^{\frac{d}{2}} \mathrm{R}_{d - 2k}(u) \; , \;\;\;
\mathrm{R}_{2k+1}(u) = - (-1)^{\frac{d}{2}} \mathrm{R}_{d - 2k-1}(u)\,.
 \ee
Taking into account (\ref{Gamma01}) we deduce
\be
 \lb{decomp7}
(\gamma_{d+1} \otimes \gamma_{d+1}) \, \frac{1}{k!} \, \gamma_{A_{k}} \otimes \gamma^{A_{k}} =
\frac{(-1)^{\frac{d}{2}}}{(d-k)!} \, \gamma_{A_{d-k}} \otimes \gamma^{A_{d-k}} \; ,
 \ee
 where $A_{d-k}$ is the multi-index such that $A_{d-k} \cap A_{k} = \emptyset$
  and $A_{d-k} \cup A_{k} = \{1,2,\dots,d\}$.
Then using (\ref{decomp6}) and (\ref{decomp7}) we immediately
obtain  (\ref{decomp5}). Equations (\ref{proj01}) follow from
(\ref{proj02}) and (\ref{decomp5}). \hfill \qed

\vspace{0.3cm}

\noindent
{\bf Proposition 2.}
The Yang-Baxter equation (\ref{YB})
is equivalent to the following relations for $\mathrm{R}^{+}$ and $\mathrm{R}^{-}$ (\ref{Rpm}):
\be \lb{prop2}
\begin{array}{c}
\mathrm{R}^{+}_{23} \, \mathrm{R}^{+}_{12} \, \mathrm{R}^{+}_{23} =
\mathrm{R}^{+}_{12}\,\mathrm{R}^{+}_{23}\,\mathrm{R}^{+}_{12}
\;\; , \;\;
\mathrm{R}^{-}_{23} \, \mathrm{R}^{+}_{12} \, \mathrm{R}^{-}_{23} =
\mathrm{R}^{-}_{12}\,\mathrm{R}^{+}_{23}\,\mathrm{R}^{-}_{12} \; ,
\\ [0.2 cm]
\mathrm{R}^{+}_{23} \, \mathrm{R}^{-}_{12} \, \mathrm{R}^{-}_{23} =
\mathrm{R}^{-}_{12}\,\mathrm{R}^{-}_{23}\,\mathrm{R}^{+}_{12}
\;\; , \;\;
\mathrm{R}^{-}_{23} \, \mathrm{R}^{-}_{12} \, \mathrm{R}^{+}_{23} =
\mathrm{R}^{+}_{12}\,\mathrm{R}^{-}_{23}\,\mathrm{R}^{-}_{12} \; ,
\\ [0.2 cm]
\mathrm{R}^{+}_{23} \, \mathrm{R}^{-}_{12} \, \mathrm{R}^{+}_{23} = 0 \, , \;\;
\mathrm{R}^{+}_{23} \, \mathrm{R}^{+}_{12} \, \mathrm{R}^{-}_{23} = 0 \, , \;\;
\mathrm{R}^{-}_{23} \, \mathrm{R}^{+}_{12} \, \mathrm{R}^{+}_{23} = 0 \, , \;\;
\mathrm{R}^{-}_{23} \, \mathrm{R}^{-}_{12} \, \mathrm{R}^{-}_{23} = 0 \, ,
\\ [0.2 cm]
\mathrm{R}^{+}_{12}\,\mathrm{R}^{-}_{23}\,\mathrm{R}^{+}_{12} = 0 \, , \;\;
\mathrm{R}^{+}_{12}\,\mathrm{R}^{+}_{23}\,\mathrm{R}^{-}_{12} = 0 \, , \;\;
\mathrm{R}^{-}_{12}\,\mathrm{R}^{+}_{23}\,\mathrm{R}^{+}_{12} = 0 \, , \;\;
\mathrm{R}^{-}_{12}\,\mathrm{R}^{-}_{23}\,\mathrm{R}^{-}_{12} = 0 \, ,
\end{array}
\ee
where the dependence on the spectral parameters is the same
as in (\ref{YB}).

\noindent
{\bf Proof.}
The Yang-Baxter equation
$\mathrm{R}^{+}_{23} \, \mathrm{R}^{+}_{12} \, \mathrm{R}^{+}_{23} =
\mathrm{R}^{+}_{12}\,\mathrm{R}^{+}_{23}\,\mathrm{R}^{+}_{12}$
is deduced from (\ref{YB}) if we act on it by projectors $\mathrm{P}^{+}_{12}$
and $\mathrm{P}^{+}_{23}$ from the left and right and use commutation relations
$$
\mathrm{P}^{+}_{12} \, \mathrm{R}^{+}_{23} =  \mathrm{R}^{+}_{23} \, \mathrm{P}^{+}_{12}  \; , \;\;
\mathrm{P}^{+}_{23} \, \mathrm{R}^{+}_{12} =  \mathrm{R}^{+}_{12} \, \mathrm{P}^{+}_{23}  \; .
$$
The relation $\mathrm{R}^{+}_{23}\,\mathrm{R}^{-}_{12}\,\mathrm{R}^{+}_{23} = 0$ is obtained as following
$$
\mathrm{R}^{+}_{23} \, \mathrm{R}^{-}_{12} \, \mathrm{R}^{+}_{23} =
\mathrm{R}^{+}_{23} \, \mathrm{R}^{-}_{12} \, \mathrm{P}^{+}_{23} \,\mathrm{R}^{+}_{23} =
\mathrm{R}^{+}_{23} \, \mathrm{P}^{-}_{23} \, \mathrm{R}^{-}_{12} \, \mathrm{R}^{+}_{23} = 0 \; ,
$$
where we use $\mathrm{R}^{-}_{12} \, \mathrm{P}^{+}_{23} = \mathrm{P}^{-}_{23} \, \mathrm{R}^{-}_{12}$, etc. \hfill \qed
\vspace{0.2cm}

We stress that in view of (\ref{prop2})  the Yang-Baxter equation (\ref{YB}) is satisfied for
any linear combination $\mathrm{R}(u)= \alpha(u) \mathrm{R}^{+}(u)+ \beta(u) \mathrm{R}^{-}(u)$
with {\it arbitrary} coefficient functions $\alpha(u)$ and $\beta(u)$. It means that
 $\mathrm{A}(u)$ and $\mathrm{B}(u)$ in (\ref{Revenodd}) are not fixed by equation (\ref{YB}).
 Moreover one can check by using (\ref{prop2}) that the Yang-Baxter equation (\ref{YB}) is satisfied if we
  transform the solution $\mathrm{R}$ as following (we write this transformation in terms of even and odd parts
 $\mathrm{R}^{+}(u)$, $\mathrm{R}^{-}(u)$):
 \be
 \lb{rep-gam0}
 \begin{array}{c}
 \mathrm{R}^{+} \to \mathrm{R}^{+} \; , \;\;\;
 \mathrm{R}^{-} \to \pm \mathrm{R}^{-} \; (\gamma_{d+1} \otimes {\bf 1}) =
 \mp (\gamma_{d+1} \otimes {\bf 1}) \; \mathrm{R}^{-}  \; , \\ [0.3cm]
 \mathrm{R}^{+} \to \mathrm{R}^{+} \; , \;\;\;
 \mathrm{R}^{-} \to \pm \mathrm{R}^{-} \; ({\bf 1} \otimes \gamma_{d+1})=
 \mp ({\bf 1} \otimes \gamma_{d+1}) \; \mathrm{R}^{-}  \; .
 \end{array}
 \ee



\noindent
{\bf Proposition 3.}
Even and odd parts  (\ref{Rpm}) of the spinorial
$\mathrm{R}$-matrix satisfy {\it unitarity} relations
\be \lb{unitarity}
 \mathrm{R}^{+}(u) \, \mathrm{R}^{+}(-u) = \mathrm{h}_{+}(u) \, \mathrm{P}^{+} \;,\;\;
 \mathrm{R}^{-}(u) \, \mathrm{R}^{-}(-u) = \mathrm{h}_{-}(u) \, \mathrm{P}^{-} \;.
\ee
where functions $\mathrm{h}_{+}(u)$, $\mathrm{h}_{-}(u)$ are constructed
out of coefficents $\mathrm{R}_k(u)$ (\ref{Revenodd})
\begin{eqnarray*}
\mathrm{h}_{+}(u) &=& 2\sum\limits_{k=0}^{d/2} \binom{d}{2k}
\mathrm{R}_{2k}(u)\, \mathrm{R}_{2k}(-u) =
\mathrm{A}(u) \mathrm{A}(-u) \prod\limits_{k=0}^{\frac{d}{2}-1}(k^2-u^2)\,, \\
\mathrm{h}_{-}(u) &=& 2\sum\limits_{k=0}^{d/2-1} \binom{d}{2k+1} \mathrm{R}_{2k+1}(u)\, \mathrm{R}_{2k+1}(-u) =
\mathrm{B}(u) \mathrm{B}(-u) \prod\limits_{k=1}^{\frac{d}{2}-1}(k^2-u^2)\,.
\end{eqnarray*}
Let us draw attention that in the right hand sides of the relations (\ref{unitarity}) projectors $\mathrm{P}^{\pm}$
(\ref{proj02}) appear.

\noindent
{\bf Proof.} At first in view of (\ref{Revenodd}) one obtains that
at special value of spectral parameter spinorial $\mathrm{R}$-matrix reduces to projector
(\ref{proj02}): $\mathrm{R}^{+}(\epsilon) = \epsilon \,\Gamma\left(\frac{d}{2}\right) \mathrm{P}^{+} + O(\epsilon^2)$ at $\epsilon \to 0$.
Then the first Yang-Baxter relation in (\ref{prop2}) at $v=-u+\epsilon$ and $\epsilon \to 0$ leads to
$\mathrm{R}^{+}_{23}(u) \, \mathrm{P}^{+}_{12} \, \mathrm{R}^{+}_{23}(-u) =
\mathrm{R}^{+}_{12}(-u) \,\mathrm{R}^{+}_{23}\,\mathrm{R}^{+}_{12}(u)$. The latter relation
is equivalent to $\mathrm{P}^{+}_{12} \, \mathrm{R}^{+}_{23}(u) \, \mathrm{R}^{+}_{23}(-u) =
\mathrm{R}^{+}_{12}(-u) \,\mathrm{R}^{+}_{12}(u) \,\mathrm{R}^{+}_{23}$ that implies
$\mathrm{R}^{+}_{12}(u) \,\mathrm{R}^{+}_{12}(-u) \sim \mathrm{P}^{+}_{12}$. In a similar manner
the second Yang-Baxter relation in (\ref{prop2}) leads to
$\mathrm{R}^{-}_{12}(u) \,\mathrm{R}^{-}_{12}(-u) \sim \mathrm{P}^{-}_{12}$. Coefficient functions $\mathrm{h}_{+}(u)$,
$\mathrm{h}_{-}(u)$ (\ref{unitarity}) are calculated in Subsection \ref{secUn} using generating function technique.
\hfill \qed
\vspace{0.2cm}

Our considerations are aimed to the
check of the Yang-Baxter equation (\ref{YB}) for the $\mathrm{R}$-matrices (\ref{r-mtr01}),
 (\ref{Revenodd}) and verification of their properties. For this we need to perform a rather
complicated computations with Clifford algebra of gamma-matrices.
To succeed in it we appeal to the technique of the generating functions developed in \cite{VDK}.
We briefly describe this technique in the next Section.

\section{Clifford algebra}
\setcounter{equation}{0}

\subsection{Fermionic interpretation of Clifford algebra}

Let $\Gamma_{a}$, $a=1,\ldots,d$, be a set of
$d$ generators of the Clifford algebra satisfying the standard relations (cf. (\ref{def}))
\begin{equation}
 \lb{def-G}
\Gamma_{a}\Gamma_{b}+\Gamma_{b}\Gamma_{a} =
2\,\delta_{a b}\, \mathbf{1}\,.
\end{equation}
The Clifford algebra is a vector
space with dimension $2^{d}$. The standard basis in this
space is formed by anti-symmetrized products of $\Gamma_a$. For example (cf. (\ref{Gamma01}))
 \begin{equation}
 \lb{bas-Cl}
\Gamma_{A_0} = \mathbf{1}\  ,\  \Gamma_{A_1} = \Gamma_{a}\ ,
\ \Gamma_{A_2} = \Gamma_{a_1 a_2} = \frac{1}{2!}\,
[\Gamma_{a_1}\Gamma_{a_2}-\Gamma_{a_2}\Gamma_{a_1}]\,,\,\cdots
 \end{equation}
 $$
 \Gamma_{A_k} = \Gamma_{a_1 \dots a_k} = {\rm As} (\Gamma_{a_1} \cdots \Gamma_{a_k} ) =
  \frac{1}{k!} \sum_s (-1)^{p(s)} \Gamma_{s(a_1)} \cdots \Gamma_{s(a_k)} \; .
 $$
 Here we use the notion of antisymmetric product $\mathrm{As}$ of $\Gamma_a$-operators.
Inside the $\mathrm{As}$-product the operators $\Gamma_a$ behave
like anti-commuting variables.

Note that the Clifford algebra generators
can be represented as~\cite{Zinn}
\be \lb{repGamma}
\Gamma_a = \theta_a + \partial_{\theta_a}\;;\;\;\;
\{ \theta_a , \theta_b \} = 0 \; ,\; \;\;
\left\{ \partial_{\theta_a} , \partial_{\theta_b} \right\} = 0 \; ,\; \;\;
\left\{ \partial_{\theta_a} , \theta_b \right\} = \delta_{a b}
\ee
where $\partial_{\theta_a}=\frac{\partial}{\partial \theta_a}$
 and $\theta_a$ $(a=1,\cdots,d)$ form a set of $d$ {\it fermionic} variables
(generators of the Gra{\ss}mann algebra).
Below the fermionic interpretation of the operators $\Gamma_a$ will be important for us
and to distinguish them from the matrices $\gamma_a$ we use different notation $\gamma_a \to \Gamma_a$.

Now we introduce the generating
function for the basis elements $\Gamma_{A_k}$ (\ref{bas-Cl})
\be \lb{genfun}
\sum_{k=0}^{\infty} \frac{1}{k!}\,u^{a_k}\cdots u^{a_1} \,\mathrm{As}\,(\Gamma_{a_1}\cdots\Gamma_{a_k}) =
\sum_{k=0}^{\infty} \frac{1}{k!}\,\left(u^{a}\Gamma_{a}\right)^k = \mathrm{exp} (u \cdot \Gamma)=
\mathrm{As}\left[ \exp(u \cdot \Gamma)\right]\,.
\ee
Here  $u \cdot \Gamma = u^a \, \Gamma_a$, $u^{a}$
are anti-commuting auxiliary variables: $u^{a}\,u^{b} = - u^{b}\,u^{a}$ and we also adopt that $u^{a}\,\Gamma_{b} = - \Gamma_{b}\,u^{a}$.
Formula (\ref{genfun}) implies that the basis elements $\Gamma_{A_k}$ (\ref{bas-Cl})
can be obtained from $\exp(u \cdot \Gamma)$ as
\be \lb{genFunGamma}
\Gamma^{a_1\ldots a_k} =
\left.\partial_{u_{a_1}}\cdots\partial_{u_{a_k}} \exp(u \cdot \Gamma)\right|_{u=0}\,.
\ee
Further we indicate two basic relations which will be used extensively in
our calculations with Clifford algebra.

\noindent
{\bf Proposition 3.} The product of generating functions (\ref{genfun}) is evaluated as
\begin{equation}\label{base}
e^{u_1 \cdot \Gamma}\cdots e^{u_k \cdot \Gamma} =
e^{-\sum_{i<j} u_i \cdot u_j}\, e^{(\sum_{i=1}^k u_i ) \cdot \Gamma}\,.
\end{equation}
 Let $u^{a}\,,\,v^{a}\,,\,\alpha^{a} \,,\,\beta^{a}$ be anti-commuting variables,
$x$ and $y$ are commuting variables. Then we have the following identity
\begin{equation}\label{use}
\Bigl. \exp\left(x \, \partial_{u} \cdot \partial_{v} \right)
\,\exp(u\cdot \alpha+v \cdot \beta+ y\,u \cdot v )\Bigr|_{u=v=0} = (1-x y)^{d}
\,\exp\left(\frac{x}{1-x y} \, \alpha \cdot \beta \right)\,,
\end{equation}
where we have used shorthand notation
$\partial_{u} \cdot \partial_{v} \equiv \frac{\partial}{\partial u^{a}}
\frac{\partial}{\partial v_{a}}$.

\noindent
{\bf Proof.} The formula (\ref{base}) is a consequence of
the Backer-Hausdorff formula
$e^A\,e^B = e^{A+B +\frac{1}{2}[A,B]} \,,$
where we take $A=u\cdot \Gamma$ , $B=v\cdot \Gamma$ and
  $[A,B] = - u^{a} v^{b}\left(\Gamma_{a}\Gamma_{b}+\Gamma_{b}\Gamma_{a}\right) = -2\,u \cdot v$.

Identity (\ref{use}) can be easily deduced by taking into account the
standard representation~\cite{V1,V2,FS,Zinn} of the operator
$\mathrm{exp}\left(x \, \partial_{u} \cdot \partial_{v}\right)$
as gaussian integral over $2d$ anti-commuting variables $\theta_a$ and $\bar{\theta}_a$. Indeed
$$
\exp\left(x \, \partial_{u} \cdot \partial_{v} \right) =
x^{d}\,\int \prod_{a=1}^d
\mathrm{d}\theta_a\,\mathrm{d}\bar{\theta}_a\,
\exp\left( x^{-1}\, \bar{\theta}\cdot\theta +
\bar{\theta} \cdot \partial_u + \partial_v \cdot \theta \right)\,,
$$
so that all operations of differentiations lead to the simple
shifts $u\to u+\bar{\theta}\,,v\to v-\theta$ and then the left
hand side of (\ref{use}) takes the form of the gaussian integral again
$$
x^{d}\, \int \prod_{a=1}^d
\mathrm{d}\theta_a\,\mathrm{d}\bar{\theta}_a\,
\exp\left( \left(x^{-1}-y\right) \bar{\theta}\cdot\theta +
\bar{\theta} \cdot \alpha + \beta \cdot \theta\right) =
x^d\,\left(x^{-1}-y\right)^d\,\exp\left(\frac{x}{1-x y} \,
\alpha \cdot \beta \right)\,.\qed
$$
In fact all subsequent calculations are based on (\ref{base}) and (\ref{use}).

Note that the topic of this section has an evident interpretation in
the language of quantum field theory.
The formula~(\ref{base}) is one of the variants of
Wick's theorem and expresses the result of reduction to
the normal form. The topic of this section
can be considered as an application of the general field-theoretical
functional technique~\cite{V1} to a very special example, and exactly this
point of view was elaborated in the paper~\cite{VDK}.
It is possible to use the language of symbols of
fermionic operators~\cite{FS} as well. For simplicity
we have derived all needed formulae in a very naive and
straightforward way.

\subsection{Fermionic realization of $\mathrm{R}$-matrix}
\label{rules}

Dealing with the Yang-Baxter equation (\ref{YB}) as well as with
$\mathrm{RLL}$-relation (\ref{RLLssf})
we have to handle the tensor product of several spinor representation spaces.
In fact we need gamma-matrices $\gamma_a$ acting in the tensor product
of two spaces. Since we consider instead of gamma-matrices $\gamma_a$
 the generators of Clifford algebra we need here two types of generators $(\Gamma_1)_{a}$, $(\Gamma_2)_{a}$
which anticommute to each other
\be \lb{grade}
  (\Gamma_1)_{a}\,(\Gamma_2)_{b} = - (\Gamma_2)_{b}\, (\Gamma_1)_{a}\, .
\ee
It is rather natural due to emphasized above fermionic nature
of representation (\ref{repGamma}).
Moreover the convention (\ref{grade}) makes the formulae much simpler.

$SO(d)$-invariant fermionic $\mathrm{R}$-matrix (\ref{r-mtr01})
is constructed out of tensor products $\left(\Gamma_1\right)_{A_k} \left(\Gamma_2\right)^{A_k}$.
Let us rewrite this gamma-matrix structure in a more appropriate form
$$
\left(\Gamma_1\right)_{A_k}  \left(\Gamma_2\right)^{A_k}= \mathrm{As}\left[\Gamma_{1a_1}\cdots\Gamma_{1a_k}\right]
\mathrm{As}\left[\Gamma_{2}^{a_1}\cdots\Gamma_{2}^{a_k}\right] =
\mathrm{As} \left[\Gamma_{1a_1}\cdots\Gamma_{1a_k}\right]
\Gamma_{2}^{a_1}\cdots\Gamma_{2}^{a_k} =
$$
\be \lb{GammaGamma}
= \mathrm{As}_{(1)} \left[\Gamma_{1a_1}\cdots\Gamma_{1a_k}\,
\Gamma_{2}^{a_1}\cdots\Gamma_{2}^{a_k}\right] =  s_k\,\mathrm{As}_{(1)} \left[(\Gamma_1\cdot\Gamma_2)^k\right]
\ee
where $s_k\equiv(-1)^{\frac{k(k-1)}{2}}$ and we denote by $\mathrm{As}_{(1)}$
the operation $\mathrm{As}$ applied only for the product of $(\Gamma_{1})_{a}$.
Below we will omit index $(1)$ in the notation $\mathrm{As}_{(1)}$ since for the expressions of the type
(\ref{GammaGamma}) we have  $\mathrm{As}_{(1)}=\mathrm{As}_{(2)}$.
At the first step in (\ref{GammaGamma}) taking into account definition (\ref{Gamma01})
one can forget about one of the symbols $\mathrm{As}$
due to convolution of two antisymmetric tensors.
Next it is possible to accomplish rearrangements taking into account
that $\Gamma_1\cdot\Gamma_2 = - \Gamma_2\cdot\Gamma_1$.
The last equality in (\ref{GammaGamma}) implies that $\mathrm{As}\left[e^{x\,\Gamma_1\cdot\Gamma_2}\right]$
is a generating function for the set of tensor products $\left(\Gamma_1\right)_{A_k}  \left(\Gamma_2\right)^{A_k}$
\be \lb{GammaGammaFun}
\mathrm{As}\left[e^{x\,\Gamma_1\cdot\Gamma_2}\right]\, = \,
\sum_k \frac{s_k}{k!} x^k \, \left(\Gamma_1\right)_{A_k}  \left(\Gamma_2\right)^{A_k}\;.
\ee
Thus we have succeeded in rewriting the multiple summation over repeated indices in a compact form.

\noindent
{\bf Proposition.}
\be \lb{asym2}
\mathrm{As}\left[e^{x\,\Gamma_1\cdot\Gamma_2}\right]=
\left.e^{x \, \partial_{u} \cdot \partial_{v}}
\,e^{u\cdot\Gamma_1+v\cdot\Gamma_2}\right|_{u=v=0}
\,,
\ee
{\bf Proof.} Using (\ref{genFunGamma}) we obtain
 \be \lb{GGG}
s_k \, \left(\Gamma_1\right)_{A_k}  \left(\Gamma_2\right)^{A_k}= \Bigl. s_k \,
\partial_{u_{a_1}} \cdots \partial_{u_{a_k}} e^{u \cdot \Gamma_1} \;
\partial_{v^{a_1}} \cdots \partial_{v^{a_k}} e^{v \cdot \Gamma_2} \Bigl|_{u=v=0}
= \Bigl.
(\partial_{u} \cdot \partial_{v})^k e^{u \cdot \Gamma_1} \; e^{v \cdot \Gamma_2} \Bigl|_{u=v=0} \; .
 \ee
 Substitution of (\ref{GGG}) into (\ref{GammaGammaFun}) gives (\ref{asym2}). \qed

 \vspace{0.2cm}

Consider a fermionic analog of the operator (\ref{r-mtr01}) where coefficient functions are assumed
to be arbitrary. Using generating function (\ref{GammaGammaFun}) we represent
this operator
in several equivalent forms
\be \lb{R}
\mathrm{R}(u) = \sum_{k=0}^{\infty} \frac{\mathrm{R}_k(u)}{k!}
\left(\Gamma_1\right)_{A_k} \left( \Gamma_2\right)^{A_k} =
\sum_{k=0}^{\infty} \frac{\mathrm{R}_k(u)\,s_k}{k!}\, \partial_{x}^k
\,\left.\mathrm{As}\left(e^{x\,\Gamma_1\cdot\Gamma_2}\right)\right|_{\lambda=0} =
\mathrm{R}(u|x)\ast\mathrm{As}\left(e^{x\,\Gamma_1\cdot\Gamma_2}\right)\,,
\ee
where we have used shorthand notation
$\mathrm{R}(x)\ast \mathrm{F}(x) \equiv
\left.\mathrm{R}(\partial_{x})\, \mathrm{F}(x)\right|_{x=0}$.
Note that all information about coefficient functions of the operator $\mathrm{R}$ in (\ref{R})
 is encoded in just one function $\mathrm{R}(u|x)$
\be \lb{Rfun}
\mathrm{R}(u|x) = \sum_{k=0}^{\infty} \frac{\mathrm{R}_k(u)\,s_k}{k!}\, x^k \,.
\ee
At the end of this Subsection we show how to represent fermionic operators (\ref{R})
 in the matrix form. There are two matrix representations $\rho'$ and $\rho^{\prime\prime}$
 for the fermionic Clifford algebra with generators
 $\Gamma_{1 a}$ and $\Gamma_{2 a}$ and defining relations (\ref{def-G}), (\ref{grade}):
 \be
 \lb{rep-gam}
 \begin{array}{c}
 \rho'(\Gamma_{1 a}) = \gamma_a \otimes {\bf 1}  \; , \;\;\;  \rho'(\Gamma_{2 a}) = \gamma_{d+1} \otimes \gamma_a
 \\ [0.3cm]
 \rho^{\prime\prime}(\Gamma_{1 a}) = \gamma_a \otimes \gamma_{d+1}  \; , \;\;\;
  \rho^{\prime\prime}(\Gamma_{2 a}) = {\bf 1} \otimes \gamma_a
 \end{array}
 \ee
 where $\gamma_a,\gamma_{d+1}$ are standard $\gamma$-matrices defined in (\ref{def}) and (\ref{chiral}).
 For the even and odd parts of (\ref{R})
 $$
 \mathrm{R}^{+} = \sum_{k=0}^{\infty} \frac{\mathrm{R}_{2k}}{(2k)!}
\left(\Gamma_1\right)_{A_{2k}} \left( \Gamma_2\right)^{A_{2k}}  \; , \;\;\;
\mathrm{R}^{-} = \sum_{k=0}^{\infty} \frac{\mathrm{R}_{2k+1}}{(2k+1)!}
\left(\Gamma_1\right)_{A_{2k+1}} \left( \Gamma_2\right)^{A_{2k+1}} \; ,
 $$
  we obtain by using (\ref{rep-gam}) the following representations
 \be
 \lb{rep-gam2}
 \rho'(\mathrm{R}^{+}) = \sum\limits_{k=0}^{\infty} \frac{\mathrm{R}_{2k}}{(2k)!}
\gamma_{A_{2k}} \otimes \gamma^{A_{2k}}  \; , \;\;\;
 \rho'(\mathrm{R}^{-}) = \left( \sum_{k=0}^{\infty} \frac{\mathrm{R}_{2k+1}}{(2k+1)!}
\gamma_{A_{2k+1}} \otimes \gamma^{A_{2k+1}} \right) (\gamma_{d+1} \otimes {\bf 1}) \; ,
\ee
 \be
 \lb{rep-gam3}
  \rho^{\prime\prime}(\mathrm{R}^{+}) = \sum\limits_{k=0}^{\infty} \frac{\mathrm{R}_{2k}}{(2k)!}
\gamma_{A_{2k}} \otimes \gamma^{A_{2k}}  \; , \;\;\;
 \rho^{\prime\prime}(\mathrm{R}^{-}) = - \left( \sum_{k=0}^{\infty} \frac{\mathrm{R}_{2k+1}}{(2k+1)!}
\gamma_{A_{2k+1}} \otimes \gamma^{A_{2k+1}} \right) ({\bf 1} \otimes \gamma_{d+1}) \; .
 \ee
 Taking into account the fact that the solutions of the Yang-Baxter equation (\ref{YB}) admit
 transformations (\ref{rep-gam0}) we can use the following convention to
 construct matrix representation $\rho$ of the Yang-Baxter solutions (\ref{R}):
 \be \lb{rho}
 \rho \left( \sum_{k=0}^{\infty} \frac{\mathrm{R}_k}{k!}
\left(\Gamma_1\right)_{A_k} \left( \Gamma_2\right)^{A_k} \right) =
\sum\limits_{k=0}^{\infty} \frac{\mathrm{R}_{k}}{k!}
\gamma_{A_{k}} \otimes \gamma^{A_{k}} \; .
 \ee
Let us note that at even $d$
\be \lb{gammagamma}
\rho'\left( \frac{1}{d!}\mathrm{As}\,(\Gamma_1 \cdot \Gamma_2)^d \right)=
\rho''\left( \frac{1}{d!}\mathrm{As}\,(\Gamma_1 \cdot \Gamma_2)^d \right)= \gamma_{d+1}\otimes\gamma_{d+1}\,.
\ee

\subsection{Exchange operators} \lb{PermOp}

In this Subsection we examine simple examples of the operators presented in the form (\ref{R}).

Let us consider the exchange operators $\mathrm{P}$, $\mathrm{P}'$
defined by means of relations
\be \lb{21}
\left(\Gamma_2\right)_a\,\mathrm{P}\, = \, \mathrm{P}\,\left(\Gamma_1\right)_{a}\;;
\;\;\;\,\left(\Gamma_1\right)_{a}\mathrm{P}' \, = \, \mathrm{P}'\left(\Gamma_2\right)_a\,.
\ee
We are going to show that it can be represented in the form (\ref{R})
\be \lb{Perm}
\mathrm{P} = e^{x}\ast\, \mathrm{As}\left(e^{x \,\Gamma_1\cdot\Gamma_2}\right)
= \mathrm{As}\left(e^{\Gamma_1\cdot\Gamma_2}\right)\;;\;\;\;
\mathrm{P}' = e^{-x}\ast\, \mathrm{As}\left(e^{x \,\Gamma_1\cdot\Gamma_2}\right)
= \mathrm{As}\left(e^{-\Gamma_1\cdot\Gamma_2}\right)\,.
\ee
To be more concrete let us rewrite previous expression for operator $\mathrm{P}$ in the following form
$$
\mathrm{P}= \sum_{k=0}^{\infty}
\frac{s_k}{k!}\, \left(\Gamma_1\right)_{A_k} \left( \Gamma_2\right)^{A_k}
= \sum_{k=0}^\infty  \left( \sum_{a_1 < a_2 < \dots < a_k}
\Gamma_{1 a_1} \Gamma_{1 a_2} \cdots \Gamma_{1 a_k} \,
\Gamma^{a_k}_2 \Gamma^{a_{k-1}}_2\cdots \Gamma^{a_1}_2 \right) \,.
$$
The proof presented below will serve as a simple example to demonstrate typical calculations
with the generating functions.
Firstly we prove identities
\be \lb{genProd}
e^{s \cdot \Gamma_1}\,\mathrm{As}\left(e^{x\, \Gamma_1 \cdot \Gamma_2}\right) =
\mathrm{As}\left(e^{x \, \Gamma_1 \cdot \Gamma_2+s \cdot (\Gamma_1+x\Gamma_{2})}\right)\ \ ;\ \
\mathrm{As}
\left(e^{x \, \Gamma_1 \cdot \Gamma_2}\right)\, e^{t \cdot \Gamma_2} =
\mathrm{As}\left(e^{x \, \Gamma_1 \cdot \Gamma_2+ t\cdot (\Gamma_2+x\Gamma_{1})}\right)\,.
\ee
The proof is rather simple and we perform it
in detail for the first product.
$$
e^{s\cdot\Gamma_1}\,\mathrm{As}\left(e^{x \, \Gamma_1\cdot\Gamma_2}\right) =
\left.
e^{x \, \partial_{u} \cdot \partial_{v}}
\,e^{s\cdot\Gamma_1}\, e^{u\cdot\Gamma_1+v\cdot\Gamma_2}
\right|_{u=v=0} =
$$
$$
= \left.
e^{x \, \partial_{u} \cdot \partial_{v}}
\,e^{u \cdot s+u \cdot \Gamma_1+v \cdot \Gamma_2+s \cdot \Gamma_1}
\right|_{u=v=0} =
 \mathrm{As}
\left(e^{x(\Gamma_1+s)\cdot \Gamma_2+s\cdot\Gamma_1}\right)\,.
$$
Here we apply successively (\ref{asym2}) and (\ref{base}). 
Thus (\ref{genProd}) is proven.
In fact this calculation set the pattern for subsequent
manipulations with generating functions.

We rewrite equation (\ref{21}) with the help of generating functions (\ref{genFunGamma}), (\ref{GammaGammaFun})
\be \lb{eqPerm}
\left.\mathrm{P}(x)\ast\,\partial_{s_a}
e^{s\cdot \Gamma_1}\,\mathrm{As}\left(e^{x \,\Gamma_1 \cdot \Gamma_2}\right) \right|_{s=0} = \left.\mathrm{P}(x)\ast\,
\partial_{t_a} \mathrm{As}
\left(e^{x \,\Gamma_1 \cdot \Gamma_2}\right)\, e^{t \cdot \Gamma_2}
\right|_{t=0}\,.
\ee
Substituting (\ref{genProd}) in (\ref{eqPerm}) and
calculating the derivatives with respect to $s_a$ and $t_a$ we obtain
$$
\mathrm{P}(x)\ast\,
\mathrm{As}\left[(\Gamma_{1a}+x\Gamma_{2a})\,e^{x \, \Gamma_1 \cdot \Gamma_2}\right] =
\mathrm{P}(x)\ast\,
\mathrm{As}
\left[(\Gamma_{2a}+x\Gamma_{1a})\,e^{x\, \Gamma_1 \cdot \Gamma_2}\right]
\,,
$$
or equivalently
\begin{equation}\label{eq1}
\left[\mathrm{P}(x)-
 \partial_x \, \mathrm{P}(x)\right]\ast\,
\mathrm{As}\left(\Gamma_{1a}\,e^{x \, \Gamma_1 \cdot \Gamma_2}\right) =
\left[\mathrm{P}(x)-
\partial_x \, \mathrm{P}(x)\right]\ast\,
\mathrm{As}
\left(\Gamma_{2a}\,e^{x\, \Gamma_1\cdot\Gamma_2}\right)\,,
\end{equation}
where in the last transformation we use the formula
$\mathrm{P}(x)\ast\,x^n\,\mathrm{F}(x) = \partial^n_{x}\mathrm{P}(x)\ast\,\mathrm{F}(x)$.
As an evident consequence of (\ref{eq1}) we obtain differential equation on the function $\mathrm{P}(x)$
$$
\partial_x \, \mathrm{P}(x)=\mathrm{P}(x) \;\; \Longrightarrow \;\;
\mathrm{P}(x) = e^{x}\,,
$$
that finishes the proof of (\ref{Perm}).

In (\ref{genProd}) we have found the product of the generating functions
(\ref{genfun}) and (\ref{GammaGammaFun}).
It is exactly what we need to check $\mathrm{RLL}$-relations (\ref{RLLssf}), (\ref{RLLf}) giving rise to condition (\ref{recurr}).
Corresponding calculation is implemented in detail in Appendix.

\subsection{Generating function for Yang-Baxter and unitarity relation} \lb{YBstr}

In this Subsection we examine thoroughly the gamma-matrix structure of the
Yang-Baxter (\ref{YB}) and unitarity (\ref{unitarity}) relations.
In order to apply technique outlined above and
in view of matrix representation $\rho$ (\ref{rho}) we consider instead their fermionic analogues,
i.e. the relations for fermionic operators (\ref{R}).

We start with fermionic version of the Yang Baxter relation (\ref{YB})
whose right hand side is a sum of operator tensor products
\be \lb{YBRHS}
\left(\Gamma_2\right)_{A_k} \left(\Gamma_3\right)^{A_k}
\left(\Gamma_1\right)_{B_k} \left(\Gamma_2\right)^{B_k}
\left(\Gamma_2\right)_{C_k} \left(\Gamma_3\right)^{C_k}
\ee
multiplied by appropriate coefficient functions of spectral parameters. According to
our approach instead of simplifying products of fermionic generators of Clifford algebra in (\ref{YBRHS})
we multiply corresponding generating functions (\ref{GammaGammaFun}) depending on parameters $x,\, y$ and $z$
\be \lb{YBRHSgen}
\mathrm{As}\left(e^{x\, \Gamma_2 \cdot\Gamma_3}\right)
\mathrm{As}\left(e^{z\, \Gamma_1 \cdot \Gamma_2}\right)
\mathrm{As}\left(e^{y\, \Gamma_2 \cdot\Gamma_3}\right) =
(1-x y)^d \mathrm{As} \left(
e^{\frac{z(1+x y)}{1-x y}\, \Gamma_1 \cdot\Gamma_2 +
\frac{x + y}{1-x y}\, \Gamma_2 \cdot \Gamma_3 + \frac{z(y-x)}{1-x y} \, \Gamma_1 \cdot\Gamma_3}
\right)\,.
\ee
Expanding the latter formula into a series over $x,\,y,\,z$ and picking out
appropriate term one obtains (\ref{YBRHS}). Let us outline derivation of (\ref{YBRHSgen}).
Using (\ref{asym2}) one can rewrite the product of the three generating functions in (\ref{YBRHSgen}) as follows
$
e^{x \partial_{u} \cdot \partial_{v}}\,
e^{y \partial_{s} \cdot \partial_{t}}\,
e^{z \partial_{p} \cdot \partial_{q}}
\,
e^{u\cdot \Gamma_2 + v\cdot \Gamma_3} \,
e^{s\cdot \Gamma_1 + t\cdot \Gamma_2} \,
e^{p\cdot \Gamma_2 + q\cdot \Gamma_3}\,.
$
Then due to (\ref{base})
$$
e^{u\cdot \Gamma_2 + v\cdot \Gamma_3} \cdot
e^{s\cdot \Gamma_1 + t\cdot \Gamma_2} \cdot
e^{p\cdot \Gamma_2 + q\cdot \Gamma_3} =
e^{s\cdot \Gamma_1 + (u+t+p)\cdot \Gamma_2 +
(v +q)\cdot \Gamma_3 + t\cdot u + p\cdot t +
q\cdot v + p\cdot u}
$$
and applying several times (\ref{use}) one obtains the desired result (\ref{YBRHSgen}).
In much the same way generating function of the
tensor product structure in the left hand side of the fermionic Yang-Baxter relation (\ref{YB}) has the form
\be \label{YBLHSgen}
\mathrm{As}\left(e^{y\, \Gamma_1 \cdot \Gamma_2}\right)
\mathrm{As}\left(e^{z\, \Gamma_2 \cdot \Gamma_3}\right)
\mathrm{As}\left(e^{x\, \Gamma_1 \cdot \Gamma_2}\right) =
(1-x y)^d \mathrm{As} \left(
e^{\frac{x + y}{1-x y}\, \Gamma_1 \cdot \Gamma_2 +
\frac{z(1+x y)}{1-x y}\, \Gamma_2 \cdot \Gamma_3 +
\frac{z(y-x)}{1-x y} \, \Gamma_1 \cdot \Gamma_3}
\right)\,.
\ee
Let us mention that expressions (\ref{YBRHSgen}) and (\ref{YBLHSgen}) are almost identical.

Dealing with unitarity relation (\ref{unitarity}) for spinorial $\mathrm{R}$-matrix
we calculate $\mathrm{R}_{12}(u)\mathrm{R}_{12}(-u)$ that forces us to consider
tensor products of fermionic generators
$
\left(\Gamma_1\right)_{A_k} \left(\Gamma_2\right)^{A_k}
\left(\Gamma_1\right)_{B_k} \left(\Gamma_2\right)^{B_k}\,.
$
Corresponding generating function is the following
\be \lb{unitGen}
\mathrm{As}\left( e^{x\,\Gamma_1\cdot\Gamma_2} \right) \mathrm{As} \left( e^{y\,\Gamma_1\cdot\Gamma_2} \right) =
(1-x y)^d \mathrm{As} \left( e^{\frac{x +y}{1-x y}\,\Gamma_1 \cdot\Gamma_2} \right)\,.
\ee
Thus we have indicated the generating functions for tensor product structure of the relevant
fermionic relations for spinorial $\mathrm{R}$-matrix (\ref{R}).
In the subsequent Sections using obtained results we will prove these relations.

\noindent
{\bf Remark.} Equations (\ref{YBRHSgen}), (\ref{YBLHSgen}), (\ref{unitGen}) give the identities
for the exchange operators (\ref{Perm}):
$$
\mathrm{P}_{12} \, \mathrm{P}_{23} \, \mathrm{P}_{12} =
 \mathrm{P}_{23} \, \mathrm{P}_{12} \, \mathrm{P}_{23} \; , \;\;\;
  \mathrm{P}'_{12} \, \mathrm{P}'_{23} \, \mathrm{P}'_{12} =
\mathrm{P}'_{23} \, \mathrm{P}'_{12} \, \mathrm{P}'_{23} \; ,
$$
$$
\mathrm{P} \, \mathrm{P} = \frac{2^d}{d\,!}\,\mathrm{As}\left(\Gamma_1\cdot\Gamma_2\right)^d\ \ \,,\ \
\mathrm{P}' \, \mathrm{P}' = \frac{(-2)^d}{d\,!}\,\mathrm{As}\left(\Gamma_1\cdot\Gamma_2\right)^d \; , \;\;\;
\mathrm{P} \, \mathrm{P}' = \mathrm{P}' \, \mathrm{P} = 2^d\,\mathbf{1} \,.
$$

\subsection{Local Yang-Baxter relation} \lb{nonloc-YB}

Let us note that due to (\ref{YBRHSgen}) and (\ref{YBLHSgen}) the
following local Yang-Baxter relation takes place
 \be
 \lb{nl-YB}
 \begin{array}{l}
(1-x y)^{-d} \,\mathrm{As}\left(e^{y \,\Gamma_1\cdot \Gamma_2}\right) \,
\mathrm{As}\left(e^{z \,\Gamma_2\cdot \Gamma_3}\right) \,
\mathrm{As}\left(e^{x \,\Gamma_1\cdot \Gamma_2}\right) = \\ [0.2cm]
=
(1-x' y')^{-d} \,
\mathrm{As}\left(e^{x' \Gamma_2\cdot \Gamma_3}\right) \,
\mathrm{As}\left(e^{z' \Gamma_1\cdot \Gamma_2}\right) \,
\mathrm{As}\left(e^{y' \Gamma_2\cdot \Gamma_3}\right) \; ,
 \end{array}
 \ee
where parameters $x,y,z$ and $x',y',z'$ are related by equations
\begin{equation}\label{Sys}
\frac{x+y}{1-xy} = \frac{z^{\prime}\,(1+x^{\prime} y^{\prime})}{1-x^{\prime} y^{\prime}}\ \,,\ \
\frac{z\,(1+x y)}{1-x y} =\frac{x^{\prime}+y^{\prime}}{1-x^{\prime} y^{\prime}}
\ \,,\ \
\frac{z\,(x-y)}{1-x y} = \frac{z^{\prime}\,(x^{\prime}-y^{\prime})}{1-x^{\prime} y^{\prime}}\,.
\end{equation}
The last relation in (\ref{Sys}) and the product of
the first two relations in (\ref{Sys}) show that the functions
$$
\lambda_1 = \frac{z\,(x-y)}{(1-x y)} \, \; , \;\;\; \lambda_2 = \frac{z\,(x+y)(1+x y)}{(1-x y)^2 }  \; ,
$$
are invariant under the transformation $x,y,z \to x',y',z'$. Thus, the points
$(x,y,z)$ and $(x^{\prime},y^{\prime},z^{\prime})$ lie on the curve $\mathcal{C}_{a,b}$ defined by the equations
\begin{align}\label{System}
\begin{cases}
z\,(x-y) = b\,(1-x y)
\\[2mm]
(x+y)(1+x y) = a\,(x-y)(1-x y)
\end{cases}
\end{align}
where $b=\lambda_1$  and $a = \frac{\lambda_2}{\lambda_1}$ are parameters which fix the curve.
The geometrical picture is the following. The second equation in (\ref{System})
defines
the family of curves parameterized by $a$ in the plane $(x,y)$. Thus, it is
possible to introduce new coordinates $(x,y) \to (a,t)$ in the plane
where $t$ is a coordinate on the curve specified by $a$.
 The variable $t$ is a coordinate on $\mathcal{C}_{a,b}$ as well.
Then due to the first equation in (\ref{System}) the coordinate
 $z$ is determined by $b$ and $(x,y)$ or equivalently by $b$ and $(a,t)$.
The transformation $(x,y,z) \to (x',y',z')$ is equivalent to the change of coordinates
$t \to t'$ on the curve $\mathcal{C}_{a,b}$.
Now we specify the coordinate $t$ on the curve
 and chose, according to (\ref{System}), new variables $(a,b,t)$ instead of $(x,y,z)$:
\be \lb{vch}
a = \frac{1+x y}{1- x y}\,\frac{x+y}{x-y} \;,\;\;\; b = z \, \frac{x-y}{1- x y}\;,\;\;\; t = \frac{x-y}{1 + x y}\,.
\ee
In terms of these new variables the transformation $x,y,z \to x',y',z'$ looks very simple
$$
 a \to a' = a \; , \;\;\; b \to b' = a \; ,
 \;\;\; t \to t' = \frac{b}{a \, t} \,.
$$
The $t \to t'$ transformation follows from the second relation in (\ref{Sys}) which can be written as $b/t = a' t'$.

At the end of this Section we note that the local Yang-Baxter equations were introduced
in \cite{Mai} and applied to the investigations of 3d integrable systems in many papers
(see e.g. \cite{Kash,Isaev}).


\section{Yang-Baxter relation and unitarity} \lb{Yang-Baxter}
\setcounter{equation}{0}

In order to prove crucial properties of the
spinorial $\mathrm{R}$-matrix (\ref{r-mtr01})
we need to transform it to a more appropriate form.
In Subsection \ref{secRint} we rewrite
spinorial $\mathrm{R}$-matrix in fermionic realization (\ref{R})
as an integral over auxiliary parameter
\be \lb{Rint}
\mathrm{R}(u) =
\int^{\infty}_{0}
\frac{\mathrm{d} x \; x^{u-1} }{(1+x^2)^{u+\frac{d}{2}}}
\left[ a(u) \, \mathrm{As}\left(e^{x \Gamma_1 \cdot\Gamma_2}\right) +
       b(u) \, \mathrm{As}\left(e^{-x \Gamma_1 \cdot\Gamma_2}\right) \right]
\ee
where $a(u)$ and $b(u)$ are two arbitrary functions related
with $\mathrm{A}(u)$ and $\mathrm{B}(u)$ appearing in (\ref{Revenodd}).
Representation (\ref{Rint}) happens to be very helpful since it enables to avoid
multiple summations over repeated indices in (\ref{r-mtr01}).
Moreover the finite summation over $k$ in (\ref{r-mtr01}) is substituted by an integral over
auxiliary parameter. Thus the Yang-Baxter equation (\ref{YB}) which would assert
equality of the two cumbersome multiple sums if we use representation (\ref{r-mtr01}),
turns into an equality of two integrals.
Using representation (\ref{Rint}) we check directly that the Yang-Baxter equation (\ref{YB})
is satisfied.
More concretely we show that the equation is equivalent
to the symmetry of a certain integral taken over the space of auxiliary parameters.

\subsection{Spinorial $\mathrm{R}$-matrix} \lb{secRint}

Previously we have shown that gamma-matrix structure of spinorial $\mathrm{R}$-matrix (\ref{r-mtr01})
can be simplified considerably using fermionic realization (\ref{R}). Now we are going to make one more step
rewriting the function $\mathrm{R}(u|x)$ in (\ref{R}) that contains all
information about coefficient functions $\mathrm{R}_k(u)$.
Let us remind that coefficient functions respect recurrence relations (\ref{recurr}).
Above we have already found their solutions (\ref{Revenodd}) containing two arbitrary functions of spectral
parameter. Using this freedom coefficient functions can be expressed in terms of Euler beta function
\begin{equation} \label{Revenodd2}
\mathrm{R}_{2k}(u) = A \, (-)^k  \,\frac{\Gamma(k+\frac{u}{2})\Gamma(\frac{u+d}{2}-k)}
{\Gamma(u+\frac{d}{2})} \;,\;\;\;
\mathrm{R}_{2k+1}(u) =  B \, (-)^k  \,\frac{\Gamma(k+\frac{u+1}{2})
\Gamma(\frac{u+d-1}{2}-k)}{\Gamma(u+\frac{d}{2})}
\end{equation}
where $A(u)$ and $B(u)$ are arbitrary functions of spectral parameter.
Then we separate even and odd terms in (\ref{Rfun})
$$
\mathrm{R}(u|y)= \sum_{k=0}^{\infty}\frac{\mathrm{R}_k(u)\, s_k}{k!}\, y^k =
\sum_{k=0}^{\infty}\frac{\mathrm{R}_{2k}(u)\, s_{2k}}{(2k)!}\, y^{2k} +
\sum_{k=0}^{\infty}\frac{\mathrm{R}_{2k+1}(u)\, s_{2k+1}}{(2k+1)!}\, y^{2k+1}\,,
$$
take into account $s_{2k} = s_{2k+1} = (-)^k$, resort to
integral representation of the $\mathrm{B}$-function
$$
\frac{\Gamma(k+\frac{u}{2})\Gamma(\frac{u+d}{2}-k)}{\Gamma(u+\frac{d}{2})} =
2 \int^{\infty}_{0} \frac{ \mathrm{d} x \; x^{u-1} \, x^{2k} }{(1+x^2)^{u+\frac{d}{2}}}
$$
$$
\frac{\Gamma(k+\frac{u+1}{2})\Gamma(\frac{u+d-1}{2}-k)}{\Gamma(u+\frac{d}{2})} =
2 \int^{\infty}_{0} \frac{ \mathrm{d} x \; x^{u-1} \, x^{2k+1} }{(1+x^2)^{u+\frac{d}{2}}}
$$
and sum up the series obtaining
\be \lb{RFun}
\mathrm{R}(u|y) =
\int^{\infty}_{0}
\frac{\mathrm{d} x \; |x|^{u-1} }{(1+x^2)^{u+\frac{d}{2}}}
\left[ (A+B) \, e^{x y} + (A-B) \, e^{-x y} \right]\,.
\ee
Thus we have managed to substitute finite set of coefficient functions appearing in (\ref{r-mtr01})
by the integral over auxiliary parameter.
Finally, applying (\ref{R}) we deduce the desired form (\ref{Rint}) of the spinorial $\mathrm{R}$-matrix
claimed above. In (\ref{decomp1}) we indicated natural decomposition of the spinorial
$\mathrm{R}$-matrix in the sum of even $\mathrm{R}^{+}$ and odd $\mathrm{R}^{-}$ parts (\ref{Rpm}).
The formulae (\ref{Rint}) and (\ref{RFun})
imply the second natural decomposition
\be \lb{decomp2}
\mathrm{R}(u) = A(u)\, \mathrm{R}^{+}(u) + B(u) \, \mathrm{R}^{-}(u) =
a(u)\, \mathcal{R}^{+}(u) + b(u) \, \mathcal{R}^{-}(u)
\ee
where
\be \lb{R1R2}
\mathcal{R}^{+}(u) \equiv
\int^{\infty}_{0}
\frac{\mathrm{d} x \; |x|^{u-1} }{(1+x^2)^{u+\frac{d}{2}}}
\, \mathrm{As}\left(e^{x \Gamma_1\cdot\Gamma_2}\right) \;,\;\;\;
\mathcal{R}^{-}(u) \equiv
\int^{\infty}_{0}
\frac{\mathrm{d} x \; |x|^{u-1} }{(1+x^2)^{u+\frac{d}{2}}} \, \mathrm{As}\left(e^{-x \Gamma_1\cdot\Gamma_2}\right)\,,
\ee
and $a= A + B,\, b = A-B$.

\subsection{Integral identity}

Now we are ready to establish the Yang-Baxter relation (\ref{YB}).
More exactly we will prove at first the Yang-Baxter relation for
spinorial $\mathrm{R}$-matrix in fermionic
realization.
Its tensor product structure has been already
discussed in Subsection \ref{YBstr}.
To be more precise corresponding generating function for
its right hand side (\ref{YBRHSgen}) and left hand side (\ref{YBLHSgen})
have been indicated.
Then in the previous Subsection we have found out that coefficient functions
of the spinorial $\mathrm{R}$-matrix can be arranged in a sole function (\ref{RFun}).
Further let us note that the Yang-Baxter relation (\ref{YB}) in fermionic realization is equivalent to the set
of eight three-term relations for $\mathcal{R}^{+},\,\mathcal{R}^{-}$ (\ref{R1R2})
\begin{equation} \label{YB2}
\mathcal{R}^{i}_{12}(u) \, \mathcal{R}^{k}_{23}(u+v) \, \mathcal{R}^{j}_{12}(v) =
\mathcal{R}^{j}_{23}(v) \, \mathcal{R}^{k}_{12}(u+v) \, \mathcal{R}^{i}_{23}(u)\,
\end{equation}
where $i,j,k = +,-$ since $a(u)$ and $b(u)$ in the expression
of spinorial $\mathrm{R}$-matrix (\ref{decomp2}) are arbitrary functions.
At first the Yang-Baxter relation will be proven for $\mathcal{R}^{+}(u)$.

Taking into account (\ref{YBRHSgen}), (\ref{YBLHSgen}) and (\ref{RFun}) one can easily see
that the Yang-Baxter relation (\ref{YB2}) at $i = j = k =+$ is equivalent to
\be \lb{YBasym}
\mathrm{As} \biggl[ \mathrm{I}^{u,v}(\Gamma_1\cdot\Gamma_2 ,
\Gamma_1\cdot\Gamma_3 , \Gamma_2\cdot\Gamma_3) \biggr]=
\mathrm{As} \biggl[ \mathrm{I}^{u,v}(\Gamma_2\cdot\Gamma_3 ,
\Gamma_1\cdot\Gamma_3 , \Gamma_1\cdot\Gamma_2) \biggr]
\ee
where
\be \lb{I}
\mathrm{I}^{u,v}(A, B , C) \equiv
\int_{D} 
\frac{ \mathrm{d} x \mathrm{d} y \mathrm{d} z  \;\;  |x|^{u-1} |y|^{v-1} |z|^{u+v-1} (1-x y)^d}
{(1+x^2)^{u+\frac{d}{2}} (1+y^2)^{v+\frac{d}{2}} (1+z^2)^{u+v+\frac{d}{2}} }
\, e^{A \,\frac{x+y}{1- x y} + B\,\frac{z(y-x)}{1- x y} +
C\, \frac{z(1+ x y)}{1 - x y}}\,,
\ee
the integration domain $D=\{ (x,y,z): x \geq 0 ,\, y \geq 0 ,\, z \geq 0 \}$.
Instead of verifying (\ref{YBasym}) we are going to check more general relation
\be \lb{II}
\mathrm{I}^{u,v}(A, B , C) =
\mathrm{I}^{u,v}(C, B , A)
\ee
where the left and right hand sides to be understood as formal power series in $A,\,B,\,C$ which are
unspecified commuting external parameters.
The discrete symmetry (\ref{II}) of the integral (\ref{I}) will be established by means of the integration variable change
$(x,y,z) \to (x',y',z')$
defined by the local Yang-Baxter equation (\ref{nl-YB}) which
leads to the system of relations (\ref{Sys}).
One can easily see that under this transformation of variables
 the external parameters $A$ and $C$  are interchanged
 in the exponential factor in (\ref{I}). However it is rather nontrivial that
the other factors in the integrand transform in the right way such that (\ref{II}) is satisfied.

To see it we appeal to geometric interpretation of the transformation (\ref{Sys})
which have been discussed in Section \ref{nonloc-YB}, where we proposed to change
variables $(x,y,z) \to (a,b,t)$ according to (\ref{vch}).
In this case the integration domain $D$ can be represented as $\bigcup_{a,b} \mathcal{C}_{a,b}$, where $\mathcal{C}_{a,b}$
is a curve parameterized by $a$ and $b$.
After all we make in (\ref{I}) the natural change of integration variables $(x,y,z)\to (a,b,t)$,
presented in (\ref{vch}), for which the Jacobian determinant has a rather simple form
$$
\left| \frac{\partial(t,a,b)}{\partial(x,y,z)} \right|= 2\,\frac{(1+x^2)(1+y^2)}{(1+ x y)(1- x y)^3}\,.
$$
\begin{figure}[tbp]
\begin{center}
\includegraphics[width = 6 cm]{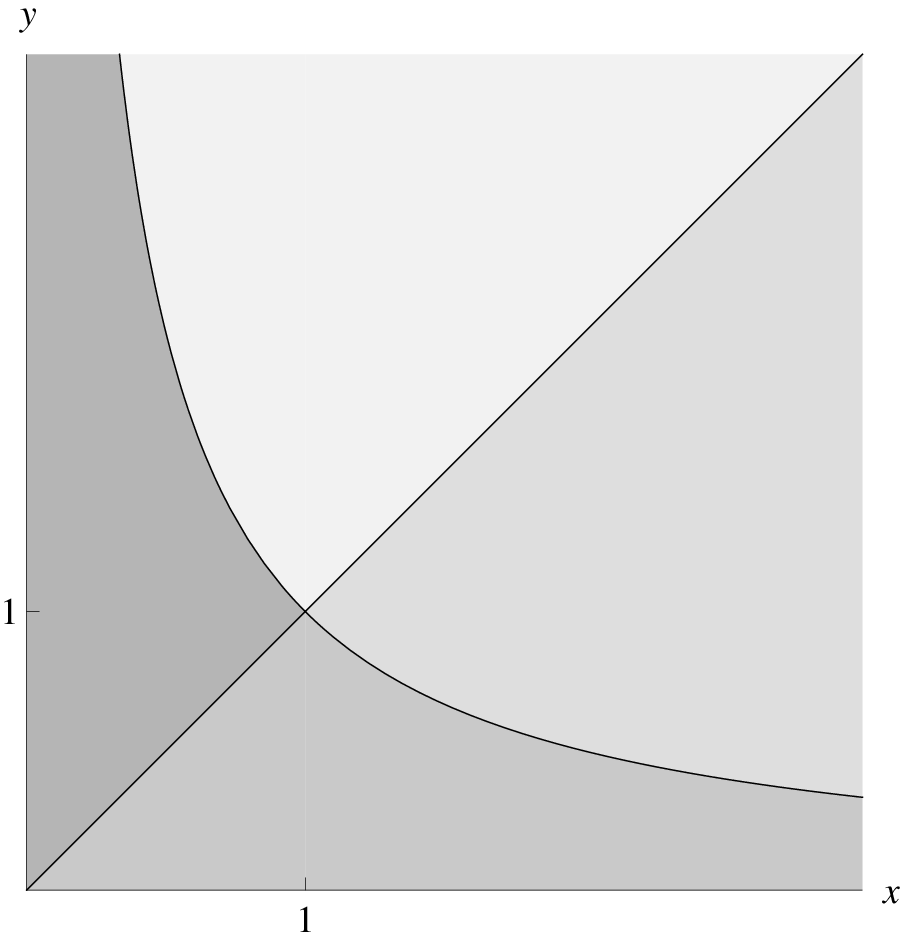} \hspace{1.5 cm}
\includegraphics[width = 6 cm]{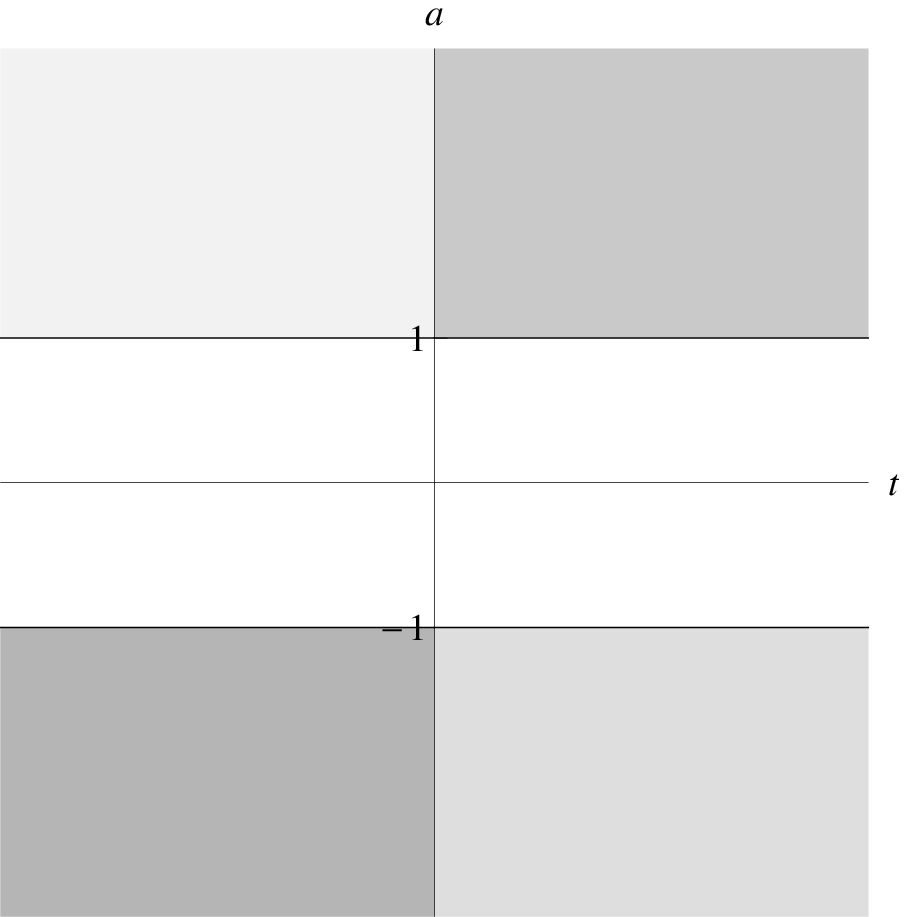}
\end{center}
\caption{\label{fig1}Projection of the domain $D$ onto the plane $(x,y)$.
It is separated by the curves $x = y,\,x y =1$ into four parts marked by different
colors, each mapped on the corresponding domain in Fig.~\ref{fig2}.
}
\caption{\label{fig2}Projection of the domain $G$ (\ref{G}) onto the plane $(t,a)$.
It is separated into four parts corresponding to four subdomains on Fig.~\ref{fig1}.}
\end{figure}
The formulae (\ref{vch}) map domain $D$ onto disconnected domain $G$
\be \lb{G}
G = \left\{(a,b,t): a\geq 1 , \, b\geq 0 \right\} \cup \left\{(a,b,t): a\leq -1 ,\, b \leq 0 \right\}
\ee
that is illustrated in Fig.~\ref{fig1},~\ref{fig2}.
After a simple calculation one obtains
\be \lb{Itab}
\mathrm{I}^{u,v}(A, B, C) =
\frac{1}{2^{u+v-1}}
\int_{G}\frac{\mathrm{d}a\,\mathrm{d}b\, \mathrm{d}t\;
|b|^{u+v-1} |t|^{2(u+v)+d-1}\exp\left(A \, a\, t -
B\, b + C \, b\, t^{-1}\right)}
{ |1+a|^{1-u} |1-a|^{1-v} \left[b^2+(1+b^2) t^2 + a^2\,t^4\right]^{u+v+\frac{d}{2}}}\,.
\ee
Since the integral (\ref{I}) is rewritten in the form (\ref{Itab}) it is straightforward to
prove the symmetry (\ref{II}) applying the integration variable change $(a,b,t) \to (a,b,t')$ where
 $t' = \frac{b}{a t}$.
It corresponds to the transposition $a \, t \rightleftarrows b \, t^{-1}$ in (\ref{Itab}).
Indeed the integrand in (\ref{Itab}) transforms correctly and the integration
domain $G$ is mapped onto itself.

Thus the Yang-Baxter relation (\ref{YB2}) at $i = j = k = +$ is established.
In a similar way the rest seven three-term relations (\ref{YB2}) can be checked.
To realize it we note that expressions (\ref{R1R2}) for $\mathcal{R}^{+}$ and $\mathcal{R}^{-}$
are almost identical. They can be obtained from each other reflecting the integration variable $x \to -x$.
In other words $\mathcal{R}^{+}$ and $\mathcal{R}^{-}$ differ solely in integration contour.
In the first case one integrates over positive semiaxis and in the second case over negative one.
Consequently to check one of the three-term relations (\ref{YB2}) we have to consider
the integral (\ref{I}) taken over appropriate reflected domain $D$.
For example at $i = j = -,\, k = +$ we integrate over $x\leq 0, y \leq 0 , z \geq 0$ in (\ref{I}).
When the variable change (\ref{vch}) is performed,
it leads to the integral (\ref{Itab}) with a certain integration domain that can be found in Fig.\ref{fig2}.
The symmetry (\ref{II}) is established as before by means of the variable change
$a \, t \rightleftarrows b \, t^{-1}$ in (\ref{Itab}) which preserves the integration domain as one can easily see.
Let us stress that algebraic manipulations needed to prove the three-term relations (\ref{YB2})
are the same in all eight cases. The only difference is in the integration domains in (\ref{I}) or (\ref{Itab}).

\begin{figure}[tbp]
\begin{center}
\includegraphics[width = 6 cm]{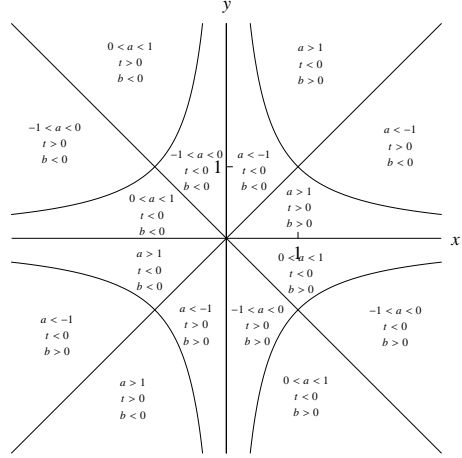}
\end{center}
\caption{\label{fig3}Projection of the domain $D$ and of its three reflections onto the plane $(x,y)$.
Images of the subregions in the space $(a,b,t)$ are indicated (see (\ref{vch})). It is assumed $z>0$.
If otherwise $z<0$ then $b$ to be substituted on the figure by $-b$.}
\end{figure}

Finally, we have checked eight three-term relations (\ref{YB2}) and hence we
have proved the Yang-Baxter relation (\ref{YB}) for
spinorial $\mathrm{R}$-matrix in fermionic realization. Using decomposition (\ref{decomp2})
of the $\mathrm{R}$-matrix in the sum of even and odd parts we obtain eight
three-term
\begin{equation} \label{YBE+-2}
\mathrm{R}^{i}_{12}(u) \, \mathrm{R}^{k}_{23}(u+v) \, \mathrm{R}^{j}_{12}(v) =
\mathrm{R}^{j}_{23}(v) \, \mathrm{R}^{k}_{12}(u+v) \, \mathrm{R}^{i}_{23}(u)\,
\end{equation}
where $i,j,k=+,-$ (compare with (\ref{prop2})).
However let us emphasize that we have always used above the fermionic representation for the $\mathrm{R}$-matrix.

At the end of Subsection~\ref{rules} we have shown how to
represent fermionic operators (\ref{R}) in the matrix form (\ref{rho}).
It can be easily checked that three-term
relations (\ref{YBE+-2}) remain valid in both matrix representation $\rho'$ and $\rho''$ (\ref{rep-gam}).
Thus the Yang-Baxter relation (\ref{YB}) for spinorial $\mathrm{R}$-matrix (\ref{r-mtr01}) is checked.


\subsection{Unitarity relation} \lb{secUn}

In the Introduction we have formulated unitarity relations (\ref{unitarity}) and proven them up to explicit
calculation of coefficient functions $\mathrm{h}_{+}(u)$, $\mathrm{h}_{-}(u)$. Now we are going to fill this gap.
We use fermionic realization of $\mathrm{R}$-matrix.
In view of (\ref{R}) and (\ref{unitGen}) one has
\be \lb{unitarity2}
\mathrm{R}^{+}(u)\mathrm{R}^{+}(-u) =
\mathrm{R}^{+}(u|x) \mathrm{R}^{+}(-u|y) \ast (1-x y)^d \mathrm{As} \left( e^{\frac{x +y}{1-x y}\,\Gamma_1 \cdot\Gamma_2} \right)\,.
\ee
Since we know that $\mathrm{R}^{+}(u)\mathrm{R}^{+}(-u)$ is proportional to projector $\mathrm{P}^{+}$,
formula (\ref{unitarity2}) contains only fermionic structures
$\mathbf{1}$ and $\mathrm{As}\left(\Gamma_1 \cdot \Gamma_2\right)^d$.
Coefficients at the other structures are equal to zero.
Thus it will be sufficient for us to calculate numerical coefficient for $\mathbf{1}$ in
(\ref{unitarity2}) that is equal to
$$
\textstyle{\mathrm{R}^{+}(u|x) \mathrm{R}^{+}(-u|y) \ast (1-x y)^d =
\sum_{k=0}^{d/2} \binom{d}{2k} \mathrm{R}_{2k}(u) \mathrm{R}_{2k}(-u)\,.}
$$
Similarly calculating numerical coefficient for $\mathrm{As}\left(\Gamma_1 \cdot \Gamma_2\right)^d$
in (\ref{unitarity2}) one obtains
$$
\textstyle{\mathrm{R}^{+}(u|x) \mathrm{R}^{+}(-u|y) \ast (x + y)^d =
\sum_{k=0}^{d/2} (-1)^{\frac{d}{2}}\binom{d}{2k} \mathrm{R}_{2k}(u) \mathrm{R}_{d-2k}(-u)\,.}
$$
Further, using matrix representations $\rho'$ (\ref{rep-gam2}) or $\rho''$ (\ref{rep-gam3})
one obtains (\ref{rho})
$$
\rho'(\mathrm{R}^{+}(u))\rho'(\mathrm{R}^{+}(-u)) =
\rho''(\mathrm{R}^{+}(u))\rho''(\mathrm{R}^{+}(-u)) = \rho(\mathrm{R}^{+}(u))\rho(\mathrm{R}^{+}(-u))\,
$$
that leads finally to the first unitarity relation (\ref{unitarity})
in view of (\ref{gammagamma}).

The previous arguments are valid also for $\mathrm{R}^{-}(u)\mathrm{R}^{-}(-u)$.
The coefficient function for $\mathbf{1}$ is equal to
$$
\textstyle{
\mathrm{R}^{-}(u|x) \mathrm{R}^{-}(-u|y) \ast (1-x y)^d =
-\sum_{k=0}^{d/2-1} \binom{d}{2k+1} \mathrm{R}_{2k+1}(u) \mathrm{R}_{2k+1}(-u)\,.}
$$
The matrix realization of the second unitarity relation (\ref{unitarity}) is provided by
$$
\rho'(\mathrm{R}^{-}(u))\rho'(\mathrm{R}^{-}(-u)) =
\rho''(\mathrm{R}^{-}(u))\rho''(\mathrm{R}^{-}(-u)) = -\rho(\mathrm{R}^{-}(u))\rho(\mathrm{R}^{-}(-u))\,.
$$

\noindent
{\bf Remark.} Unitarity relations (\ref{unitarity}) can be established as well be means of
integral representation for $\mathrm{R}$-matrix (\ref{Rint}) using
$$
\int\limits^{\infty}_{0} \int\limits^{\infty}_{0}
\frac{\mathrm{d} x \,\mathrm{d} y \; x^{u-1} y^{-u-1}
                                (x+y)^k(1 - x y)^{d-k}}
     {(1+x^2)^{u+\frac{d}{2}} (1+y^2)^{-u+\frac{d}{2}}} =
\left\{
\begin{array}{ccl}
-\frac{2 \pi}{u}\frac{1}{ \sin \pi u} \;,\; & \text{at} & \; k = 0 \\ [0.2 cm]
-\frac{2 \pi}{u}\cot \pi u \;,\; & \text{at} & \; k = d \\ [0.2 cm]
0 \;,\; & \text{at} & \; k =1,\cdots,d-1\,.
\end{array}
\right.
$$

\section*{Acknowledgment}


The work of D.C. is supported by the Chebyshev Laboratory
(Department of Mathematics and Mechanics, St.-Petersburg State University)
under RF government grant 11.G34.31.0026,
and by Dmitry Zimin's "Dynasty" Foundation.
The work of S.D. is partially supported by RFBR grants
11-01-00570-a and 12-02-91052.
The work of A.P.Isaev was supported by the grant RFBR 11-01-00980-a
and grant Higher School of Economics No.11-09-0038.





\appendix
\renewcommand{\theequation}{\Alph{section}.\arabic{equation}}
\setcounter{table}{0}
\renewcommand{\thetable}{\Alph{table}}



\section{Appendix}
\setcounter{equation}{0}

In \cite{CDI} we introduced representation space $V'$ which we assumed to be infinite-dimensional in general
and restricted the universal Yang-Baxter equation (\ref{RRR}) to the space $V \otimes V \otimes V'$
\begin{equation}
 \label{RLLf}
\mathrm{R}_{12}(u - v) \, \mathrm{L}_{13}(u) \,  \mathrm{L}_{23}(v) =
\mathrm{L}_{13}(v) \, \mathrm{L}_{23}(u)  \,  \mathrm{R}_{12}(u - v)
   \;\; \in  \;\; {\rm End}(V \otimes V \otimes V') \; .
\end{equation}
The operator $\mathrm{L}(u)$ which is defined in the tensor product
$V \otimes V'$ of spinor and arbitrary representation $T'$ spaces
has been sought for in the form
\begin{equation}
\label{RLLf-2}
\mathrm{L}(u) = u+\frac{i}{4}\,\gamma_{a b}\otimes T'(M^{a b})\,.
\end{equation}
Here notation (\ref{bas-Cl}) is used and $M_{ab}$ $(a,b = 1,\dots, d)$ are generators of $so(d)$
subjected to relations
\begin{equation} \lb{so}
[ M_{ab} , M_{dc} ] = i( \delta_{bd} M_{ac}
+ \delta_{ac} M_{bd} - \delta_{ad} M_{bc} - \delta_{bc} M_{ad} )\,.
\end{equation}
In \cite{CDI} we claimed that $\mathrm{RLL}$-relation (\ref{RLLf}) with spinorial $\mathrm{R}$-matrix (\ref{r-mtr01})
is satisfied if representation $T'$ is such that
\begin{equation}\label{asym}
T' \left( \left\{M_{[ab}\,,M_{c]d}\right\} \right) = 0 \; ,
\end{equation}
where $\left\{A\,,B\right\} = A\, B + B \, A$ is anticommutator and
square brackets denote antisymmetrization.
We will undertake corresponding calculation in the first part of this
Appendix using generating function technique.
We are going to show that $\mathrm{RLL}$-relation (\ref{RLLf})
with $\mathrm{L}$-operator (\ref{RLLf-2}) and $\mathrm{R}$-matrix of the form
(\ref{r-mtr01}) leads to recurrence relation (\ref{recurr}) for coefficient functions $\mathrm{R}_k(u)$ and
set up restriction (\ref{asym}) on the representation $T'$ in the quantum space.

In order to avoid misunderstandings let us note that in a special case $d = 6$ we have
the isomorphism $so(6,\mathbb{C}) = sl(4,\mathbb{C})$,
corresponding $8$-dimensional $\mathrm{L}$-operator (\ref{RLLf-2}) is a direct
sum of two $4$-dimensional $\mathrm{L}$-operators of $s\ell(4)$ algebra,
spinorial $\mathrm{R}$-matrix (\ref{r-mtr01})
reduces to Yang $\mathrm{R}$-matrix under Weyl projections and condition (\ref{asym}) on representation
$T'$ happens to be superfluous. We demonstrate it in the second part of this Appendix.


\subsection{$\mathrm{RLL}$-relation}

Further by abuse of notation we denote
$T'(M_{ab}) \rightarrow M_{ab}$. The following calculation is very similar to the one presented in Subsection \ref{PermOp},
and it uses the generating function technique.
We are going to prove
fermionic version of $\mathrm{RLL}$-relation (\ref{RLLf}).
Then taking matrix representation $\rho'$ or $\rho''$ (\ref{rep-gam}) one obtains
immediately (\ref{RLLf}) for spinorial $\mathrm{R}$-matrix (\ref{r-mtr01}) and $\mathrm{L}$-operator (\ref{RLLf-2}).

The substitution of
spinorial $\mathrm{R}$-matrix (\ref{R}) in fermionic realization
and fermionic analogue of $\mathrm{L}$-operator (\ref{RLLf-2}) with
unspecified representation $T'$ in the quantum space
 in $\mathrm{RLL}$-relation (\ref{RLLf}) gives
$$
\sum_{k=0}^{\infty} \frac{\mathrm{R}_k(u-v)}{k!}
\left(\Gamma_1\right)_{A_k} \left(\Gamma_2\right)^{A_k}\, \left(u+\frac{i}{4}\left(\Gamma_1\right)_{a b} M^{a b}\right)
\left(v+\frac{i}{4}\left(\Gamma_2\right)_{c d} M^{c d}\right) =
$$
\be \lb{RLL1}
=\sum_{k=0}^{\infty} \frac{\mathrm{R}_k(u-v)}{k!} \left(v+\frac{i}{4}
\left(\Gamma_1\right)_{a b} M^{a b}\right)
\left(u+\frac{i}{4} \left(\Gamma_2\right)_{c d} M^{c d}\right)
\left(\Gamma_1\right)_{A_k} \left(\Gamma_2\right)^{A_k}\,.
\ee
This relation contains terms linear and quadratic in generators
$M_{a b}$.
The product of two generators can be transformed by means of Lie algebra commutation relations (\ref{so})
$$
M_{a b}\,M_{c d} =
\frac{1}{2}\,\left[M_{a b}\,,M_{c d}\right]+
\frac{1}{2}\,\left\{M_{a b}\,,M_{c d}\right\}
= \frac{i}{2}\,\left[g_{b c}M_{a d}-g_{a d}M_{c b}-
g_{a c}M_{b d}+
g_{b d}M_{c a}\right]+
\frac{1}{2}\,
\left\{M_{a b}\,,M_{c d}\right\}
$$
so that
$$
\left(\Gamma_1\right)_{a b} \left(\Gamma_2\right)_{c d}\,
M^{a b}\,M^{c d} =
-2 i\left(\Gamma_1\right)_{a}^{\,\,\, c}\left(\Gamma_2\right)_{b c}\, M^{a b}
+\frac{1}{2}\left(\Gamma_1\right)_{a b} \left(\Gamma_2\right)_{c d}
\left\{M^{a b}\,,M^{c d}\right\}\,.
$$
All terms in (\ref{RLL1}) linear on spectral parameters
are combined in a single one $\sim(u-v)$ due to relation
$$
\left(\Gamma_1\right)_{A_k} \left(\Gamma_2\right)^{A_k}\left(\Gamma_2\right)_{a b} -
\left(\Gamma_1\right)_{a b}\left(\Gamma_1\right)_{A_k} \left(\Gamma_2\right)^{A_k} =
\left(\Gamma_1\right)_{A_k}\left(\Gamma_2\right)_{a b}\left(\Gamma_2\right)^{A_k} -
\left(\Gamma_1\right)_{A_k}\left(\Gamma_1\right)_{a b} \left(\Gamma_2\right)^{A_K}\,,
$$
that is a consequence of the $so(d)$ invariance
$$
\left[\,
\left(\Gamma_1\right)_{a b} + \left(\Gamma_2\right)_{a b}\,,
\left(\Gamma_1\right)_{A_k} \left(\Gamma_2\right)^{A_k}\,\right] = 0\,.
$$
After all
intertwining relation (\ref{RLL1}) is reduced to the form
$$
\sum_{k=0}^{\infty} \frac{\mathrm{R}_k(u)}{k!}\, u\, M^{a b}\,\biggl( \left(\Gamma_1\right)_{A_k} \left(\Gamma_2\right)^{A_k}
\left(\Gamma_2\right)_{a b} -
\left(\Gamma_1\right)_{a b}\left(\Gamma_1\right)_{A_k}\left(\Gamma_2\right)^{A_k}\biggl) +
$$
$$
+\frac{1}{2}\,\sum_{k=0}^{\infty} \frac{\mathrm{R}_k(u)}{k!}\, M^{a b}\,
\biggl( \left(\Gamma_1\right)_{A_k}\left(\Gamma_1\right)_{a}^{\,\,\, c}\left(\Gamma_2\right)^{A_k}\left(\Gamma_2\right)_{b c}-
\left(\Gamma_1\right)_{a}^{\,\,\, c}\left(\Gamma_1\right)_{A_k}\left(\Gamma_2\right)_{b c}\left(\Gamma_2\right)^{A_k}
\biggl) +
$$
\be \lb{RLL2}
+\frac{i}{8}\,\sum_{k=0}^{\infty} \frac{\mathrm{R}_k(u)}{k!}\,
\biggl( \left(\Gamma_1\right)_{A_k}\left(\Gamma_1\right)_{a b}\left(\Gamma_2\right)^{A_k}\left(\Gamma_2\right)_{c d}-
\left(\Gamma_1\right)_{c d}\left(\Gamma_1\right)_{A_k}\left(\Gamma_2\right)_{a b}\left(\Gamma_2\right)^{A_k}\biggl)
\,
\left\{M^{a b}\,,M^{c d}\right\} = 0\,.
\ee
Using the reference formulae for products of generating functions (see (\ref{genProd}))
$$
\mathrm{As}\left(e^{x \Gamma_1 \cdot \Gamma_2}\right)
\, e^{s \cdot \Gamma_1} = \mathrm{As}\left(e^{x \Gamma_1 \cdot \Gamma_2+s \cdot (\Gamma_1-x \Gamma_{2})}\right)\ \ ;\ \
e^{s \cdot \Gamma_1}\,\mathrm{As}\left(e^{x \Gamma_1 \cdot \Gamma_2}\right) = \mathrm{As}\left(e^{x \Gamma_1 \cdot \Gamma_2+s\cdot(\Gamma_1+x \Gamma_{2})}\right)\,,
$$
$$
\mathrm{As}
\left(e^{x \Gamma_1 \cdot \Gamma_2}\right)\, e^{t \cdot \Gamma_2} =
\mathrm{As}\left(e^{x \Gamma_1 \cdot \Gamma_2+t \cdot (\Gamma_2+x \Gamma_{1})}\right)\ \ ;\ \ e^{t \cdot \Gamma_2}\,\mathrm{As}
\left(e^{x \Gamma_1 \cdot \Gamma_2}\right) =
\mathrm{As}\left(e^{x \Gamma_1 \cdot \Gamma_2+t \cdot(\Gamma_2-x \Gamma_{1})}\right)\,,
$$
it is easy to derive compact expression for the first term in (\ref{RLL2})
$$
\sum_{k=0}^{\infty} \frac{\mathrm{R}_k}{k!} M^{a b}
\left[ \left(\Gamma_1\right)_{A_k} \left(\Gamma_2\right)^{A_k}\left(\Gamma_2\right)_{a b} -
\left(\Gamma_1\right)_{a b}\left(\Gamma_1\right)_{A_k}\left(\Gamma_2\right)^{A_k}\right] =
$$
$$
=
\mathrm{R}(x)\ast
M^{a b} \partial_{s_{a}} \partial_{s_{b}}
\mathrm{As}\,e^{x \Gamma_1 \cdot \Gamma_2}\left[e^{s \cdot (\Gamma_2+x \Gamma_{1})} - e^{s \cdot (\Gamma_1+x \Gamma_{2})}\right] =
$$
$$
= \mathrm{R}(x)\ast\,
M^{a b}
\,\mathrm{As}\,e^{x \Gamma_1 \cdot \Gamma_2}\,
\biggl[(\Gamma_{2a}+x \Gamma_{1b})(\Gamma_{2b}+x \Gamma_{1a}) - (\Gamma_{1a}+x \Gamma_{2a})
(\Gamma_{1b}+x \Gamma_{2b})\biggr] =
$$
$$
= \mathrm{R}(x)\ast\left(x^2-1\right)
M^{a b}
\mathrm{As}\,e^{x \Gamma_1 \cdot \Gamma_2}
\left[\Gamma_{1a}\Gamma_{1b}-
\Gamma_{2a}\Gamma_{2b}\right] =  \left(\partial^2_x\mathrm{R}(x) - \mathrm{R}(x)\right)\ast
M^{a b}
\mathrm{As}\,
e^{x \Gamma_1 \cdot \Gamma_2}\left[
\Gamma_{1a}\Gamma_{1b}-\Gamma_{2a}\Gamma_{2b}\right]\,.
$$
In a similar way using
$$
\mathrm{As}\left(e^{x \Gamma_1 \cdot \Gamma_2}\right) \, e^{s \cdot \Gamma_1+t \cdot \Gamma_2} =
\mathrm{As}\left(e^{x (\Gamma_1-s) \cdot (\Gamma_2-t)+s\cdot\Gamma_1+t\cdot\Gamma_{2}}\right)\,,
$$
$$
e^{s \cdot \Gamma_1+t \cdot \Gamma_2}\,\mathrm{As}\left(e^{x \Gamma_1 \cdot \Gamma_2}\right) =
\mathrm{As}\left(e^{x (\Gamma_1+s) \cdot (\Gamma_2+t)+s \cdot \Gamma_1+t \cdot \Gamma_{2}}\right)\,,
$$
the second term in (\ref{RLL2}) can be rearranged as follows
$$
\sum_{k=0}^{\infty} \frac{\mathrm{R}_k}{k!}\,M^{a b}\,
\left( \left(\Gamma_1\right)_{A_k}\left(\Gamma_1\right)_{a}^{\,\,\, c}\left(\Gamma_2\right)^{A_k}\left(\Gamma_2\right)_{b c}-
\left(\Gamma_1\right)_{a}^{\,\,\, c}\left(\Gamma_1\right)_{A_k}\left(\Gamma_2\right)_{b c}\left(\Gamma_2\right)^{A_k}\right) =
$$
$$
= -2\,\mathrm{R}(x)\ast\,
M^{a b}\,
\,\mathrm{As}\,e^{x \Gamma_1 \cdot \Gamma_2}\,
\left[\Gamma_{1a}\Gamma_{1b}-
\Gamma_{2a}\Gamma_{2b}\right]
\biggl[(x^3+x)\,\Gamma_{1c}\Gamma_2^{c}-(d-2)\,x^2
\biggl] =
$$
$$
= -2\biggl[x\,\partial^3_{x}\mathrm{R}(x) + x\,\partial_{x}\mathrm{R}(x)-(d-2)\,\partial^2_{x}\mathrm{R}(x)\biggl]\,
M^{a b}\,
\,\mathrm{As}\,
e^{x \Gamma_1 \cdot \Gamma_2}\,\left[\Gamma_{1a}\Gamma_{1b}-
\Gamma_{2a}\Gamma_{2b}\right]\,,
$$
and the last term in (\ref{RLL2}) takes the form
$$
\sum_{k=0}^{\infty} \frac{\mathrm{R}_k}{k!}\,
\biggl( \left(\Gamma_1\right)_{A_k}\left(\Gamma_1\right)_{a b}\left(\Gamma_2\right)^{A_k}\left(\Gamma_2\right)_{c d}-
\left(\Gamma_1\right)_{c d}\left(\Gamma_1\right)_{A_k}\left(\Gamma_2\right)_{a b}\left(\Gamma_2\right)^{A_k}\biggl)
\left\{M^{a b}\,,M^{c d}\right\} =
$$
$$
= 4\,\mathrm{R}(x)\ast\,(x^3-x)\,
\left\{M^{a b}\,,M^{c d}\right\}
\mathrm{As}\,e^{x \Gamma_1\cdot\Gamma_2}\,
\biggl[\Gamma_{1a}\Gamma_{1b}\Gamma_{1c}\Gamma_{2d} - \Gamma_{2a}\Gamma_{2b}\Gamma_{2c}\Gamma_{1d}
\biggl] =
$$
$$
= 4\,\biggl[\partial^3_{x}\mathrm{R}(x)-\partial_{x}\mathrm{R}(x)\biggl]\ast\,
\left\{M^{a b}\,,M^{c d}\right\}
\mathrm{As}\,e^{x \Gamma_1 \cdot \Gamma_2}\,
\biggl[\Gamma_{1a}\Gamma_{1b}\Gamma_{1c}\Gamma_{2d} - \Gamma_{2a}\Gamma_{2b}\Gamma_{2c}\Gamma_{1d}
\biggl]\,.
$$
Thus finally we obtain that (\ref{RLL2}) is equivalent to the relation
$$
\biggl[x\,\partial^3_{x}\mathrm{R}(x) + x\,\partial_{x}\mathrm{R}(x)
- (d-2)\,\partial^2_{x}\mathrm{R}(x)
-u \left( \partial^2_{x}\mathrm{R}(x)-\mathrm{R}(x)\right)\biggl]\ast\,
M^{a b}\,
\,\mathrm{As}\,
e^{x \Gamma_1 \cdot \Gamma_2}\,\left[\Gamma_{1a}\Gamma_{1b}-
\Gamma_{2a}\Gamma_{2b}\right] -
$$
\be \lb{RLL3}
-\frac{i}{2}\,\biggl[\partial^3_{x}\mathrm{R}(x)-\partial_{x}\mathrm{R}(x)\biggl]\ast\,
\left\{M^{a b}\,,M^{c d}\right\}
\mathrm{As}\,e^{x \Gamma_1 \cdot \Gamma_2}\,
\biggl[\Gamma_{1a}\Gamma_{1b}\Gamma_{1c}\Gamma_{2d} - \Gamma_{2a}\Gamma_{2b}\Gamma_{2c}\Gamma_{1d}
\biggl] = 0\,.
\ee
There are two independent gamma-matrix structures in the latter formula so that the
differential equation for the coefficient function $\mathrm{R}(x)$
$$
x\biggl[\partial^3_{x}\mathrm{R}(x) +
 \partial_{x}\mathrm{R}(x)\biggl] - (d-2)\,\partial^2_{x}\mathrm{R}(x)
   -u \biggl[\partial^2_{x}\mathrm{R}^{\prime\prime}(x)-\mathrm{R}(x)\biggl] = 0
$$
and requirement $\left\{M_{[ a b}\,,M_{c ] d}\right\} = 0$ (see (\ref{asym})) arise.
The differential equation produces the recurrence
relation (see (\ref{recurr})) for the
coefficients $\mathrm{R}_k(u)$:
$$
\mathrm{R}(x) = \sum_{k=0}^{\infty}\frac{s_k\,\mathrm{R}_k(u)}{k!}\,
x^k\ \longrightarrow \mathrm{R}_{k+2}(u) = - \frac{u+k}{u+d-2-k}\,
\mathrm{R}_{k}(u)\,.
$$

\subsection{$\mathrm{R}$-matrix in the special case $d=6$}

Now we proceed to the special case $d=6$.
The recurrent relations (\ref{recurr}) for odd and even coefficients are independent that
enables us to fix $\mathrm{R}_{0}(u) = (u+4)/8$ and $\mathrm{R}_1(u) = 0$.
Hence $\mathrm{R}$-matrix (\ref{r-mtr01}) takes the form
\begin{equation} \label{Rn=4}
\mathrm{R}(u) =
\mathrm{R}_0(u) \, \mathbf{1} \otimes \mathbf{1} +
\frac{\mathrm{R}_2(u)}{2!} \, \gamma_{a_1 a_2} \otimes
 \gamma^{a_1 a_2} +
\frac{\mathrm{R}_4(u)}{4!} \, \gamma_{a_1 \ldots a_4} \otimes
 \gamma^{a_1 \ldots a_4}
+ \frac{\mathrm{R}_6(u)}{6!} \, \gamma_{a_1 \ldots a_6} \otimes
 \gamma^{a_1 \ldots a_6}
\end{equation}
where
$$\mathrm{R}_0 (u) = (u+4)/8 \; , \; \mathrm{R}_2 (u) = - u/8 \; , \; \mathrm{R}_4 (u) =  u/8 \; , \; \mathrm{R}_6 (u) = - (u + 4)/8$$
and
the last term in (\ref{RLL3}) which is responsible for the condition (\ref{asym}) reduces to
\begin{equation} \label{acmn=4}
\frac{2}{3!}\biggl[
\mathrm{R}_{6}(u)+\mathrm{R}_{4}(u)\biggl]\,
\left\{M^{ab}\,,M^{c}{}_{d}\right\}\,
\biggl[\gamma_{ab c c_1 c_2 c_3}\otimes\gamma^{d c_1 c_2 c_3} + \gamma^{d c_1 c_2 c_3}\otimes
\gamma_{a b c c_1 c_2 c_3}\biggl]\;.
\end{equation}
All the other terms vanish because of the special form of coefficients $\mathrm{R}_k (u)$ and
owing to finiteness of the Clifford algebra of gamma-matrices.
Next we note that owing to
$\alpha \,\gamma_{a b c c_1 c_2 c_3} = \epsilon_{a b c c_1 c_2 c_3} \,\gamma_7$
and $\gamma_7 \gamma_7 = \mathbf{1}$ (\ref{chiral})
the gamma-matrix structure in (\ref{acmn=4}) can be transformed as follows
$$
\gamma_{a b c c_1 c_2 c_3}\otimes\gamma^{d c_1 c_2 c_3} = \gamma_7 \otimes \gamma_7 \,\gamma_{a b c c_1 c_2 c_3} \gamma^{d c_1 c_2 c_3} =
120 \,\gamma_7 \otimes \gamma_7  \left[ \, \delta^{d}_{a} \, \gamma_{b c} - \delta^{d}_{b} \, \gamma_{a c} + \delta^{d}_{c} \, \gamma_{a b} \, \right]\;.
$$
Consequently (\ref{acmn=4}) which is proportional to
$$
\left\{M^{ab}\,,M^{c}{}_{d}\right\} \left[ \, \delta^{d}_{a} \, \gamma_{b c} - \delta^{d}_{b} \, \gamma_{a c} + \delta^{d}_{c} \, \gamma_{a b} \,\right] = 2 \left\{M^{a (b}\,,M^{c)}{}_{a}\right\} \, \gamma_{b c} = 0
$$
turns to zero. In the last expression the parentheses $(...)$ denote symmetrization.
Therefore $\mathrm{RLL}$-equation (\ref{RLL1}) is valid for
arbitrary representation of generators $\{M_{ab}\}$ of the algebra $so(6)$.

Let us rewrite the expression for $\mathrm{R}$-matrix (\ref{Rn=4}) in a more transparent form.
All gamma-matrix structures in (\ref{Rn=4})
have block-diagonal form in Weyl representation for gamma-matrices. Therefore it is reasonable to consider
projections of (\ref{Rn=4}) on corresponding irreducible subspaces.
We introduce subspaces $V_{+}$ and $V_{-}$ obtained by Weyl projections:
 $V_+ = \frac{1+\Gamma_7}{2} V$ and $V_- =\frac{1-\Gamma_7}{2}V$.
At first we note that relations
$$
\left[ \mathbf{1} \otimes \mathbf{1} -
\frac{1}{6!} \, \gamma_{A_6} \otimes \gamma^{A_6} \right]_{V_{+}\otimes V_{-}} =
\left[ \frac{1}{2!} \, \gamma_{A_2} \otimes \gamma^{A_2} -
\frac{1}{4!} \, \gamma_{A_4} \otimes \gamma^{A_4} \right]_{V_{+}\otimes V_{-}} = 0\;
$$
lead to
$\left.\mathrm{R}(u)\right|_{V_{+}\otimes V_{-}} =
\left.\mathrm{R}(u)\right|_{V_{-}\otimes V_{+}} = 0$.
Further a pair of relations
$$
\left[ \mathbf{1} \otimes \mathbf{1} +
\frac{1}{6!} \, \gamma_{A_6} \otimes \gamma^{A_6} \right]_{V_{-}\otimes V_{-}} =
\left[ \frac{1}{2!} \, \gamma_{A_2} \otimes \gamma^{A_2} +
\frac{1}{4!} \, \gamma_{A_4} \otimes \gamma^{A_4} \right]_{V_{-}\otimes V_{-}} = 0
$$
leads to Yang $\mathrm{R}$-matrix
$$
\left.\mathrm{R}(u)\right|_{V_{-}\otimes V_{-}} = \left[ 2 \;\mathrm{R}_0 (u) \, \mathbf{1} \otimes \mathbf{1} +
\mathrm{R}_2 (u) \, \gamma_{a b} \otimes \gamma^{a b} \right]_{V_{-}\otimes V_{-}} =
{\bf 1} \otimes {\bf 1} + u \, \mathrm{P}\;
$$
where $\mathrm{P}$ is a permutation operator and we take into account
$
\left. - \frac{1}{8} \, \gamma_{ab} \otimes \gamma^{ab} \right|_{ V_{-}\otimes V_{-}} =
\mathrm{P} - \frac{1}{4} \, {\bf 1} \otimes {\bf 1}\,.
$
Analogously one concludes that $\left.\mathrm{R}(u)\right|_{V_{+}\otimes V_{+}} = {\bf 1} \otimes {\bf 1} + u \, \mathrm{P}\;$.


\end{document}